\documentclass[fleqn,usenatbib]{mnras}

\usepackage{newtxtext,newtxmath}

\usepackage[T1]{fontenc}
\usepackage{ae,aecompl}

\usepackage{graphicx}	
\usepackage{amsmath}	
\usepackage{amssymb}	
\usepackage{multirow}
\usepackage{longtable}
\usepackage{supertabular,booktabs}
\usepackage{float}
\usepackage{ragged2e}
\usepackage{pdflscape}
\usepackage{afterpage}
\usepackage{subcaption}
\usepackage{capt-of}
\usepackage{rotating}

\title[SNR 1E 0102.2-7219 X-ray spectra]{X-ray spectral analysis of the neutron star in SNR 1E 0102.2-7219}

\author[Hebbar, Heinke \& Ho]{
Pavan R Hebbar,$^{1}$\thanks{E-mail: hebbar@ualberta.ca}
Craig O. Heinke,$^{1}$
Wynn C. G. Ho$^{2, 3}$
\\
$^{1}$Dept. of Physics, CCIS 4-183, University of Alberta, Edmonton, AB, T6G 2E1, Canada\\
$^{2}$Department of Physics and Astronomy, Haverford College, 370 Lancaster Avenue, Haverford, PA 19041, USA\\
$^{3}$Mathematical Sciences, Physics and Astronomy and STAG Research Centre, University of Southampton, Southampton SO17 1BJ, UK
}

\date{Accepted XXX. Received YYY; in original form ZZZ}

\pubyear{2018}

\begin{document}
\label{firstpage}
\pagerange{\pageref{firstpage}--\pageref{lastpage}}
\maketitle

\begin{abstract}
We re-analysed numerous archival {\it Chandra} X-ray observations of the bright supernova remnant (SNR) 1E 0102.2-7219 in the Small Magellanic Cloud, to validate the detection of a neutron star (NS) in the SNR by \citet{vogt2018}. Careful attention to the background is necessary in this spectral analysis. We find that a blackbody + power-law model is a decent fit, suggestive of a relatively strong $B$ field and synchrotron radiation, as in a normal young pulsar, though the thermal luminosity would be unusually high for young pulsars. Among realistic NS atmosphere models, a carbon atmosphere with $B = 10^{12}$G  best fits the observed X-ray spectra. Comparing its unusually high thermal luminosity ($L_{bol} = 1.1_{-0.5}^{+1.6}\times10^{34}$ ergs s$^{-1}$) to other NSs, we find that its luminosity can be explained by decay of an initially strong magnetic field  (as in magnetars or high B-field pulsars) or by slower cooling after the supernova explosion. The nature of the NS in this SNR (and of others in the Magellanic Clouds) could be nicely confirmed by an X-ray telescope with angular resolution like {\it Chandra}, but superior spectral resolution and effective area, such as the {\it Lynx} concept.
\end{abstract}




\section{Introduction}
\label{sec:intro}

The detection of the Crab \citep{staelin1968} and Vela pulsars \citep{large1968} within supernova remnants (SNRs) verified the theory that neutron stars (NSs) are produced in these supernova explosions \citep{baade1934}. However, not all likely core-collapse SNRs contain pulsars; some contain NSs in other manifestations, while some show no known compact object at their centres. Failures to detect NSs in deep X-ray surveys of nearby Galactic SNRs suggest that these SNRs produced black holes, exceptionally cold young NSs, or no compact remnant at all \citep{kaplan2004,kaplan2006}. Thus, searching for and identifying NSs in young SNRs is essential to understand supernovae in more detail.
 
X-ray studies have been one of the most effective means to find NSs in SNRs, as NSs can generate bright X-ray emission through magnetosphere processes, or simply by re-radiating heat retained since their formation (which does not require that they be radio pulsars). X-ray observations by {\it Chandra} and {\it XMM-Newton} have helped reveal that NSs in young SNRs form a very diverse population. Of order 1/3 of young core-collapse SNRs have proposed evidence of associated NSs \citep{kaspi2002}. Considering selection effects, it is plausible that a majority of core-collapse SNRs may contain radio pulsars. A significant number of young SNRs, however, contain central X-ray sources that do not show any signs of radio pulsar activity. These include anomalous X-ray pulsars (AXPs) and soft gamma ray repeaters (SGRs), now understood to be manifestations of high-$B$ ($B \sim 10^{14}$--$10^{15}$ G) NSs known as magnetars \citep{thompson1995,thompson1996,woods2006}. At least nine magnetars are now confidently associated with supernova remnants, roughly 1/3 of known magnetars \citep{Olausen14,Gavriil08,Rea16,D'Ai16}. The central X-ray sources in SNRs also include central compact objects (CCOs), showing purely thermal (blackbody-like) X-ray emission without radio or gamma-ray counterparts \citep{pavlov2004a}.  Three CCOs with detectable pulsations now have estimated $B$ fields (from spindown) in the range $3\times10^{10}<B<10^{11}$ G \citep{gotthelf13}. Nine to eleven CCOs are now known in SNRs \citep[e.g.][]{gotthelf13,Klochkov16}.

Rapidly rotating pulsars produce energetic charged particles, typically detectable in both radio and X-ray via synchrotron emission, as pulsed emission and/or as an extended pulsar wind nebula \citep{Gaensler06,Kaspi06,Li08}. The X-ray spectra of young radio pulsars ($\tau \lessapprox 10^3$ yr) generally include a primary non-thermal component, along with blackbody (BB)-like X-ray emission from the surface, often dominated by the hotter parts of the surface near the magnetic poles \citep[e.g.][]{pavlov2001a,deLuca05,Manzali07}. These spectra generally have temperatures between 40 and 200 eV when fit by BB models, and inferred radii between 1 and 10 km \citep[e.g.][]{Page04}. For old pulsars ($\tau \gtrapprox 10^6$ yr), the NS surface becomes cool, and the X-ray emission is primarily due to thermal radiation from heated magnetic polar caps, with weak non-thermal components \citep[e.g. PSR J0437-4715,][]{Zavlin02,Bogdanov06}. CCO X-ray spectra can be fit by exclusively thermal spectra, with blackbody temperatures of a few hundred eV \citep{pavlov2002}, while magnetars have more complex spectra (see below), that may be parametrised (below 10 keV) as hot (0.3-0.5 keV) BB plus a hard power-law contribution \citep[e.g.][]{Kaspi17}.

The thermal radiation of the NS is significantly affected by its gravitational mass, radii, magnetic field, surface temperature and the composition of the NS atmosphere \citep[see][for a  review]{Potekhin14a}. Thus modelling the effects of these parameters on the NS X-ray spectra are essential to learn the properties of the NSs. Strong magnetic fields increase the binding energy of atoms and molecules \citep[e.g.][]{lai2001},  affecting the thermal radiation from the NS atmosphere and lead to cyclotron resonance scattering that mimics a separate power-law component \citep{Lyutikov06}. The assumption of a hydrogen atmosphere should be valid for most NSs where fall-back or accretion has occurred, since the elements stratify quickly to leave the lightest (generally H) on top \citep{Alcock80,romani1987,brown1998}. Thermonuclear burning of light elements on the hot young NS surface may remove H and He, possibly leaving a C (or higher-Z) atmosphere, if fallback and accretion are kept to very low rates \citep{chang2004,chang2010,Wijngaarden19}. The heavy elements in such a mid-Z atmosphere can lead to detectable spectral features \citep{ho09,mori2007}. If such features can be confidently identified, the gravitational redshift of these spectral features would be a crucial constraint on the NS mass and radius, and thus on the dense matter equation of state.

\begin{figure*}
    \centering
    \includegraphics[width=0.9\textwidth]{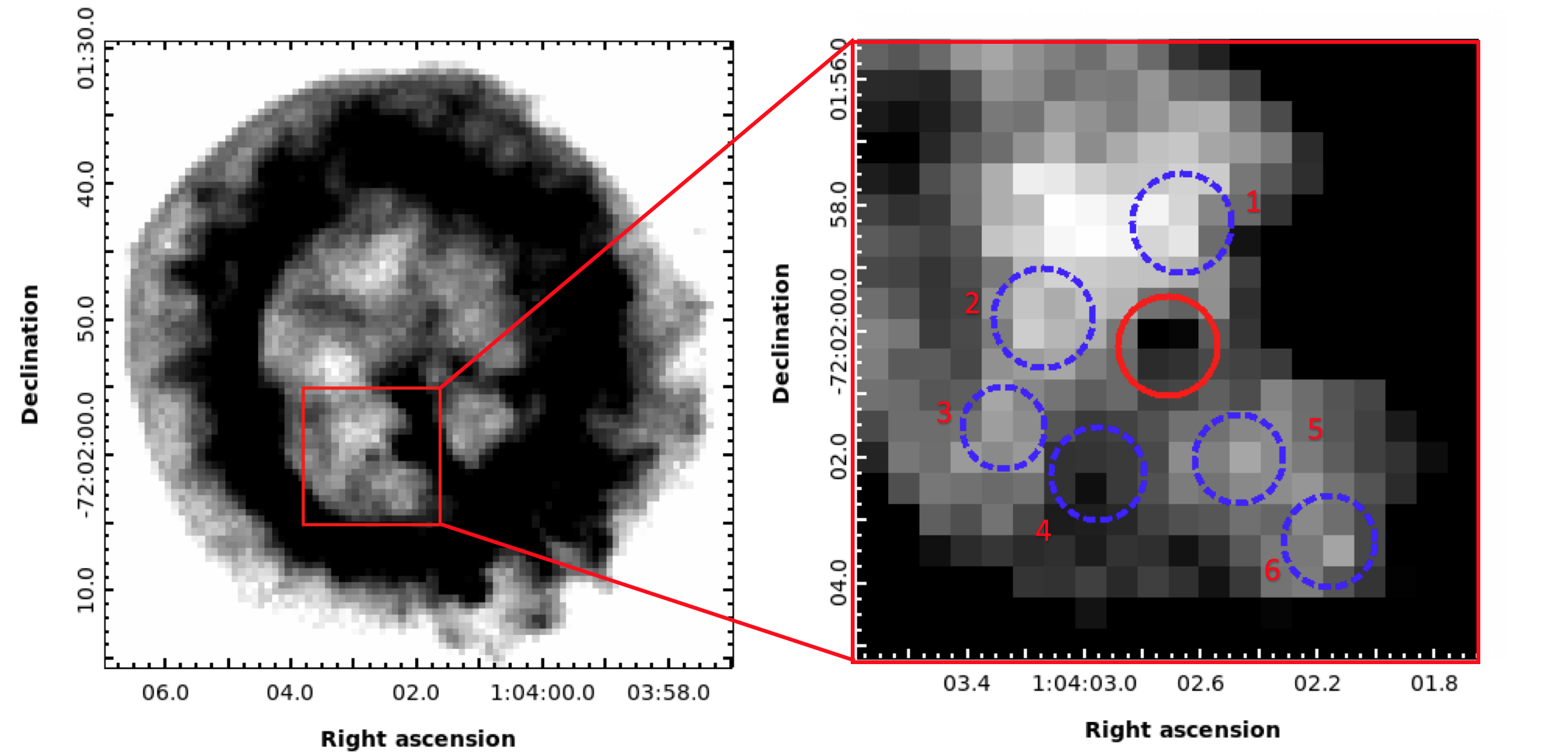}
    \caption{({\it Left}) X-ray image of SNR 1E 0102.2-7219. ({\it Right}) Magnified image showing the position of the candidate NS within the SNR. The solid red circle (radius $1 \arcsec$) is our source extraction region. The dashed blue labelled circles denote background regions used in our study (see Section~\ref{sec:back}). Background emission from the SNR contributes significantly to the flux from the source region.}
    \label{fig:regions}
\end{figure*}

\subsection{SNR 1E 0102.2-7219}
The SNR 1E 0102.2-7219 (hereafter, E0102), discovered by the {\it Einstein} observatory \citep{seward1981}, is the second brightest X-ray source in the Small Magellanic Cloud (SMC, we assume a distance of 62 kpc to SMC). The high X-ray flux from E0102 and its strong emission lines of O, Ne and Mg make it an ideal calibration source in soft X-rays for instruments aboard {\it Chandra, Suzaku, Swift} and {\it XMM-Newton} \citep{plucinsky2017}. Optical analysis of this SNR and the filaments in its ejecta have revealed the SNR to be an oxygen-rich \citep{dopita1981,tuohy1983} SNR with an age of $2050 \pm 600$ years \citep{finkelstein2006, Xi_19}. \citet{blair2000} suggested it to have been produced by a Type Ib supernova, based on the non-detection of emission from O burning products (S, Ca, Ar). However, the recent detections of [S II], [S III], [Ar III] and crucially H$\alpha$ by \cite{seitenzahl2018} in the fast-moving ejecta of the SNR provide support for a Type IIb nature. Regardless, E0102 is certain to be a SNR formed by the core collapse of a massive star, and thus is expected to have left a compact object. 

Recently, \citet{vogt2018} reported the detection of a compact object in E0102, from {\it Chandra} X-ray observations \citep{rutkowski_2010, Xi_19}, supported by MUSE identification of a low-ionisation nebula surrounding the X-ray point source. However, they did not report a direct, detailed spectral analysis of the X-ray point source, in part due to the complexity of the X-ray background in this region. Density and temperature variations within the SNR can make it difficult to subtract the background directly, leading to large residuals in the background-subtracted spectrum. However, the analysis by \citet{vogt2018}, using only 4 spectral bins and only rough comparison between simple models and the X-ray data, does not allow discrimination between different possible models for the X-ray emission from NSs.

In this paper, we re-analysed the X-ray data used by \citet{vogt2018}, to verify the presence of a NS, and study the properties of this NS. In Section~\ref{sec: obs}, we describe our data reduction methods and how we extracted the X-ray spectra. The details of the spectral analysis -- the comparison between different models and the effects of background, and the search for pulsations are discussed in Section~\ref{sec:results}. In Section~\ref{sec:disc}, we discuss the implications of these results on the properties of the neutron star.

\begin{table}
\caption{List of observations used for our analysis. All observations were taken with the ACIS-S aimpoint in the TE VFAINT mode.}
\label{table:obs_list}
\begin{tabular}{c c c c}
\hline
{\bf ObsID} & {\bf Exposure(ks)} & {\bf Start time} \\
\hline
3519 & 8.01 & 2003-02-01 04:35:57 \\
3520 & 7.63 & 2003-02-01 07:09:04 \\
3544 & 7.86 & 2003-08-10 16:08:11	\\
3545 & 7.86 & 2003-08-08 15:18:06 \\
5130 & 19.41 & 2004-04-09 13:07:53 \\
5131 &	8.01 &	2004-04-05 4:48:28 \\
6042 &	18.9 &	2005-04-12 1:40:38 \\
6043 &	7.85 &	2005-04-18 8:43:17 \\
6075 &	7.85 &	2004-12-18 1:22:38 \\
6758 &	8.06 &	2006-03-19 4:29:32 \\
6759 &	17.91 &	2006-03-21 23:29:49 \\
6765 &	7.64 &	2006-03-19 18:52:07 \\
6766 &	19.7 &	2006-06-06 13:25:56 \\
8365 &	20.98 &	2007-02-11 17:16:39 \\
9694 &	19.2 &	2008-02-07 8:05:46 \\
10654 &	7.31 &	2009-03-01 0:05:34 \\
10655 &	6.81 &	2009-03-01 2:25:44 \\
10656 &	7.76 &	2009-03-06 9:34:35 \\
10657 &	7.64 &	2009-03-06 12:09:44 \\
11957 & 18.45 &	2009-12-30 7:49:44 \\
13093 & 19.05 &	2011-02-01 2:40:10 \\
14258 &	19.05 &	2012-01-12 5:07:31 \\
15467 &	19.08 &	2013-01-28 16:33:52 \\
16589 &	9.57 &	2014-03-27 9:49:15 \\
18418 &	14.33 &	2016-03-15 16:54:42 \\
19850 &	14.33 &	2017-03-19 2:27:32 \\ \hline
\end{tabular} 

\end{table}

\section{Observations and Data Reduction}
\label{sec: obs}

The supernova remnant E0102 has been extensively observed by {\it Chandra}, and used as a calibration source to model the response of the CCD instruments \citep{plucinsky2017,alan2018}. However, the background X-ray flux from E0102 around the compact object reported by \citet{vogt2018} is extremely strong, necessitating the high angular resolution of the {\it Chandra X-ray Observatory}. We used {\it Chandra} ACIS-S VFAINT observations taken in  timed exposure mode, and pointed within $1.2 \arcmin$ of SNR 1E 0102.2-7219 ($\alpha = 01^{\mathrm{h}} 04^{\mathrm{m}} 02 \fs 4; \delta = -72^{\circ} 01 \arcmin 55 \farcs 3$). We excluded observations with signs of background flaring. We provide the list of all observations used for our analysis in Table~\ref{table:obs_list}. We point the readers to \citet{plucinsky2017,vogt2018,alan2018} for detailed comments on these observations.

We reprocessed all the data according to CALDB 4.7.6 standards using the command \texttt{chandra\_repro} in CIAO 4.10 \citep{fruscione2006}. We extracted events from a $1 \arcsec$ region around the point source ($\alpha = 01^{\mathrm{h}} 04^{\mathrm{m}} 02 \fs 7; \delta = -72^{\circ} 02 \arcmin 00 \farcs 2$) as a compromise between encompassing as large a fraction as possible of the point-spread function (PSF) of the {\it Chandra} ACIS-S instrument, and reducing the contribution from the background. This region captures about 90\% of energy from a point source at $\sim 1.5$ keV. The source and the background regions used for the analysis are shown in Fig~\ref{fig:regions}. These region files were shifted manually for individual observations to account for shifts in the astrometry. 

We extracted the spectra from each observation separately using the CIAO tool \texttt{specextract}, which considers the PSF and the encircled energy fractions while constructing the effective area files for the spectral analysis.
The increasing contamination of the ACIS detectors has resulted in degradation of their low-energy quantum efficiency over the years\footnote{http://cxc.harvard.edu/ciao/why/acisqecontamN0010.html} \citep{marshall2004}. Thus, individual observations have different effective areas and response matrix files. Assuming a linear response of the instruments, the spectra from these observations can be combined using  \texttt{combine\_spectra}. 
The \texttt{combine\_spectra} tool of CIAO 4.10 also generates exposure weighted response matrix and ancillary response files (RMFs and ARFs respectively)  for proper spectral analysis \footnote{http://cxc.harvard.edu/ciao/ahelp/combine\_spectra.html}. 
The combined spectra have more counts and better statistics,  allowing us to resolve narrow lines and test the quality (``goodness") of different models. 
We grouped the X-ray photons of the combined spectra so that each spectral bin has a width of at least 50 eV ($\sim 0.5$ times the spectral resolution of {\it Chandra} ACIS) and contains at least 15 photons. 
As uncertainties in the bright background radiation from the supernova remnant itself dominate our uncertainties, we modelled the source spectra along with the background, instead of subtracting the background. 
We only considered the interval $0.5$--$10.0$ keV, where the {\it Chandra} ACIS-S instrument is most responsive. We used XSPEC v12.9.1m for  spectral analysis. We adopted C-statistics\footnote{https://heasarc.gsfc.nasa.gov/xanadu/xspec/manual/node304.html} \citep{cash1979} for spectral fitting since it has been shown to be relatively unbiased, compared to $\chi^2$ statistics, to fit Poissonian data \citep{humphrey2009}.

We also verified our primary spectral results from the methods above by  modelling all the spectra simultaneously, rather than combining them and modelling the combined spectrum. For this purpose, we combined the spectra from observations in a given year (within each year, the responses change only slightly) using \texttt{combine\_spectra} and loaded each of these separately into XSPEC. We then modelled these spectra simultaneously by linking the corresponding parameters for each data set. Grouping the spectra to 15 photons per bin leads to large bins and would miss narrow emission features. Therefore, we grouped the spectra such that each bin contains a minimum of 1 photon,  and used C-statistics for our analysis. 

In both cases, we compared the quality of different models using the Akaike Information Criterion \citep{akaike1974} with correction for small sample sizes \citep[AICc; see][]{cavanaugh1997}.
\begin{equation}
\mathrm{AICc} = 2k + \mathrm{cstat}_{min} + \frac{2k^2 + 2k}{n-k-1},
\end{equation}
where $k$ is the number of parameters and ${\rm cstat}_{\rm min}$ is the negative of twice the logarithm of the likelihood of the model, with best-fitting parameters assuming Poisson statistics. A smaller AICc value indicates a better likelihood. The quantity $\exp$((AICc$_i$ - AICc$_j$)/2) gives the relative likelihood of model $j$ with respect to model $i$ (smaller AICc values are better fits). However, AICc cannot comment on the absolute quality of the fit of a given model. For this purpose, we used the $\chi^2$ test for the combined spectrum, and the ``goodness" simulations of XSPEC, using the Cramer-Von Mises (CvM) statistic \citep{cramer1928} for individual spectra analysed simultaneously. XSPEC's ``goodness" simulations report the fraction of simulations having a CvM statistic smaller than that for the observed spectrum. A large value of this fraction (e.g. 99\%) indicates that the observed spectrum is unlikely to be produced from the given model, and hence the model can be rejected (with 99\% confidence).

\section{Results}
\label{sec:results}
Given the high background flux, we use a detailed approach to model the background and the source spectra. In Section~\ref{sec:combined_an}, we perform our primary spectral analysis on the combined spectra of E0102. In Section~\ref{sec:sim_an}, we check the role of changing ACIS-S response on our results by loading spectra from each year separately and simultaneously modelling them.  In Section~\ref{sec:back}, we also verify our results by using different background models for our analysis. We search for pulsations in E0102 in Section~\ref{sec:pulse}.

\subsection{Analysis of combined spectra}
\label{sec:combined_an}
\begin{figure}
    \centering
    \includegraphics[width = 0.45\textwidth]{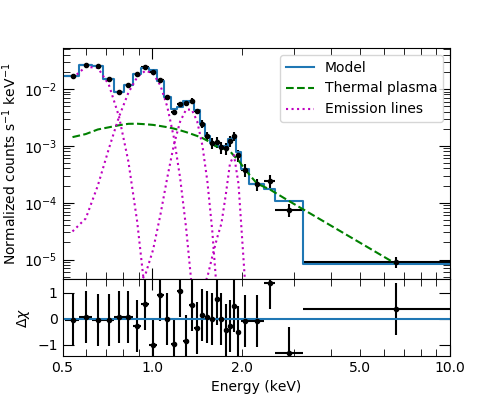}
    \caption{Empirical spectral fit to the background spectrum (data and model in top panel, residuals to the fit in units of the data uncertainties below). Our model consists of a thermal plasma with $kT = 1.2 \pm 0.3$ keV and gaussian lines at $E_1 = 0.60 \pm 0.01$ keV, $E_2 = 0.95 \pm 0.01$ keV, $E_3 = 1.32_{-0.01}^{+0.02}$ keV \& $E_4 = 1.88 \pm 0.03$ keV, corresponding to the emission lines of O, Ne, Mg and Si respectively. With $\chi^2$/d.o.f = 10.64/16, the model is a good fit to the observed background spectrum.}
    \label{fig:bkg_bkg}
\end{figure}

\begin{figure}
    \centering
    \includegraphics[width=0.45\textwidth]{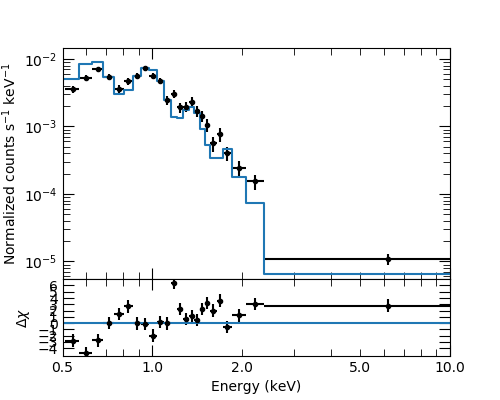}
    \caption{Source spectrum without background subtraction, fit with the scaled background model alone. The $\chi^2$-test gives $\chi^2$/d.o.f = 132.17/19, resulting in a null hypothesis probability of $p \sim 10^{-16}$. The large residuals require an additional component along with the background.
    }
    \label{fig:src_bkg}
\end{figure}

\begin{table}
    \centering
    \caption{Background model used for the spectral analysis of the compact object in SNR 1E 0102.2-7219.}
    \begin{tabular}{c c}
         \hline
         \multirow{2}{*}[0.0cm]{\bf Parameters} & \multirow{2}{*}[0.0cm]{\bf Parameter Values} \\
         & \\
         \hline
         $N_{H, MW}$ & $5.36 \times 10^{20}$ cm$^{-2}$ \\[0.1cm]
         $N_{H, SMC}$ & $5.76 \times 10^{20}$ cm$^{-2}$ \\[0.1cm]
         $kT$ & $1.2 \pm 0.3$\\[0.1cm]
         Flux$_{apec}$ & $1.5_{-0.3}^{+0.4} \times 10^{-14}$ ergs cm$^{-2}$ s$^{-1}$ \\[0.1cm]
         $E_{O}$ & $0.60 \pm 0.01$ keV \\[0.1cm]
         $\sigma_O$ & $0.065_{-0.007}^{+0.008}$ keV \\[0.1cm]
         Flux$_O$ & $4.4_{-0.3}^{+0.2} \times 10^{-14}$ ergs cm$^{-2}$ s$^{-1}$ \\[0.1cm]
         $E_{Ne}$ & $0.95 \pm 0.01$ keV \\[0.1cm]
         $\sigma_{Ne}$ & $0.084_{-0.007}^{+0.008}$ keV \\[0.1cm]
         Flux$_{Ne}$ & $2.1_{-0.2}^{+0.1} \times 10^{-14}$ ergs cm$^{-2}$ s$^{-1}$ \\[0.1cm]
         $E_{Mg}$ & $1.32_{-0.01}^{+0.02} keV$ \\[0.1cm]
         $\sigma_{Mg}$ & $0.07 \pm 0.02$ \\[0.1cm]
         Flux$_{Mg}$ & $3.7_{-0.7}^{+0.8} \times 10^{-15}$ ergs cm$^{-2}$ s$^{-1}$ \\[0.1cm]
         $E_{Si}$ & $1.88 \pm 0.03 keV$ \\[0.1cm]
         $\sigma_{Si}$ & $<0.07$ \\[0.1cm]
         Flux$_{Si}$ & $5_{-2}^{+3} \times 10^{-16}$ ergs cm$^{-2}$ s$^{-1}$ \\[0.1cm] \hline
        \end{tabular} \\
    \justify
    Note: Our background model is \texttt{wabs*tbabs*(apec + gaussian + gaussian + gaussian + gaussian)}. We use $N_{H, MW}$ and $N_{H, SMC}$ values reported in \citet{plucinsky2017} for our analysis. The \texttt{tbabs} component utilises the SMC values given in \citet{russell1992} to model intrinsic absorption in the SMC. We report the unabsorbed flux in each of these emission lines between 0.5 and 10 keV\\
    $^{*}$ Lower error bound reaches the lower hard limit. \newline
    $^{**}$ Upper error bound exceeds the upper hard limit. \newline
    \label{tab:bkg_spec}
\end{table}

We first confirm the presence of a compact object in the supernova remnant. \citet{plucinsky2017} and \citet{alan2018} do not consider regions near the candidate compact object in their spectral analysis. We analysed spectra derived from six different nearby background regions, and find that all six regions have similar spectra. Therefore we use these regions together to model the background flux. We fit the background using an empirical model consisting of a hot collisionally ionised plasma model (\texttt{apec}) with no lines (this was achieved in XSPEC using the command \texttt{xset APECNOLINES yes}) for the continuum and four Gaussian lines (\texttt{gaussian}) to model the emission line complexes from O, Ne, Mg and Si. We used two components for absorption --- \texttt{wabs} (which uses the \citet{anders1982} abundances) for absorption within the Milky Way, and \texttt{tbabs} with SMC abundances \citep{russell1992} for absorption by gas in the SMC. (We note that since the absorption within the Milky Way towards the SMC is small, using \texttt{wabs} instead of \texttt{tbvarabs} with \citet{wilms2000} parameters doesn't change our results significantly.) We fixed the absorption column of the Milky Way in this direction to $5.36 \times 10^{20}$ cm$^{-2}$, and that of the SMC to $5.76 \times 10^{20}$ cm$^{-2}$ \citep{plucinsky2017} (allowing these to vary doesn't change the C-statistics or $\chi^2$ significantly, as they are small). Thus our background model was \texttt{wabs * tbabs * (apec + gaussian + gaussian + gaussian + gaussian)}. 

Fitting this model gave $kT = 1.2 \pm 0.3$ keV and gaussian lines at $E_1 = 0.60 \pm 0.01$ keV (O complex, $\sigma = 0.065_{-0.007}^{+0.008}$ keV), $E_2 = 0.95 \pm 0.01$ keV (Ne complex, $\sigma = 0.084_{-0.007}^{+0.008}$ keV), $E_3 = 1.32 \pm 0.02$ keV (Mg complex, $\sigma = 0.07 \pm 0.02$) keV) \& $E_4 = 1.88 \pm 0.03$ keV (Si line, $\sigma < 0.07$ keV) and a C-statistic value of 10.64 for 16 degrees of freedom (d.o.f). Using this simple empirical model, we found $\chi^2$/d.o.f $= 10.64/16$. This corresponds to a null hypothesis probability\footnote{The null hypothesis probability indicates the fraction of simulated data sets drawn from the model that would have larger (worse) statistics than the real data; thus, $p=0.01$ would indicate only a 1\% probability of obtaining such a poor statistic by chance, and thus that the real data is probably not drawn from the model.} of $p=0.87$, indicating that our background model is a good fit to the observed background spectrum as shown in Fig.~\ref{fig:bkg_bkg}. We also check the fits to individual background regions using this model and find that all the parameters are similar within their error limits.

In order to check if the source emission can also be explained as only a spatial concentration of the same emission as the background, we fit this background model to the source spectrum (without background subtraction), permitting scaling of the background normalisation. We also allow for slight changes in the central energies of the emission lines, which could be due to different radial velocities. Fitting our background model alone to the source spectrum gives C-statistic/d.o.f. $= 132.17/19$ ($\chi^2$ = 119.15). This is a very poor fit ($p \sim 10^{-16}$) (shown in Fig.~\ref{fig:src_bkg}), indicating that additional components are required.

\begin{table*}
\centering
\caption{Summary of modelling the combined spectrum of the compact object in SNR 1E 0102.2-7219}
\label{table:spec_e0102}
\begin{tabular}{c  c  c  c  c c}
\hline
\multirow{2}{*}[0.0cm]{\bf Model}  &	\multirow{2}{*}[0.0cm]{\bf Parameters} & \multirow{2}{*}[0cm]{\bf Parameter Values} &	\multirow{2}{*}[0.0cm]{ \bf C-statistic} & \multirow{2}{*}[0.0cm]{\bf AICc} & {\bf $\mathbf{\chi^2/d.o.f}$}  \\
& & & & & {\bf (p-value)}\\
\hline
\multirow{4}{1.25in}[0.0cm]{\centering \textsc{wabs*tbabs*bbodyrad}} & $f$ &	$0.21_{-0.04}^{+0.03}$ & \multirow{4}{0.8in}[0.0cm]{\centering $34.89$} & \multirow{4}{0.8in}[0.0cm]{\centering $60.49$} & \multirow{4}{0.8in}[0.0cm]{\centering $32.41/16$ \\ ($0.0088$)}\\[0.1cm]
    & $N_{H, SMC}$ & $2_{-2}^{+8*} \times 10^{21}$ cm$^{-2}$ & & &\\[0.1cm]
    & $T_{BB}$ & $(2.9 \pm 0.6) \times 10^6$ K & & &\\[0.1cm]
    & $R_{BB}$ & $3_{-1}^{+4}$ km & & & \\[0.1cm] \hline
 \multirow{6}{1.3in}[-0.2cm]{\centering \textsc{wabs*tbabs*(bbodyrad + pegpwrlw)}} & $f$ & $0.19 \pm 0.04$ & \multirow{6}{0.8in}[0.0cm]{\centering $18.11$} & \multirow{6}{0.8in}[0.0cm]{\centering $48.97$} & \multirow{6}{0.8in}[0.0cm]{\centering $17.33/15$ \\ ($0.30$)} \\[0.1cm]
    & $N_{H, SMC}$ & $5.6_{-5.1}^{+10.7} \times 10^{21}$ cm$^{-2}$ & & & \\[0.1cm]
    & $T_{BB}$ & $ 2.1_{-0.8}^{+0.4} \times 10^6$ K & & & \\[0.1cm]
    & $R_{BB}$ & $8_{-4}^{+7**}$ km & & & \\[0.1cm]
	& $\Gamma$ &	$2.0$ & & & \\[0.1cm]
    & Flux$_{pl}$ & $(3 \pm 1) \times 10^{-15}$ ergs cm$^{-2}$ s$^{-1}$ & & & \\[0.1cm] \hline
\multirow{6}{1.3in}[-0.2cm]{\centering \textsc{wabs*tbabs*(bbodyrad + bbodyrad)}} & $f$ & $0.20 \pm 0.03$ & \multirow{6}{0.8in}[0.0cm]{\centering $16.72$} & \multirow{6}{0.8in}[0.0cm]{\centering $53.64$} & \multirow{6}{0.8in}[0.0cm]{\centering $16.30/14$ \\ ($0.30$)} \\[0.1cm]
    & $N_{H, SMC}$ & $1.0_{-0.9}^{+0.6} \times 10^{22}$ cm$^{-2}$ & & & \\[0.1cm]
    & $T_{BB, c}$ & $ 1.8_{-0.1}^{+0.6} \times 10^6$ K & & & \\[0.1cm]
    & $R_{BB, c}$ & $15_{-10}^{+0.0**}$ km & & & \\[0.1cm]
	& $T_{BB, h}$ & $ 6_{-1}^{+8} \times 10^6$ K & & & \\[0.1cm]
    & $R_{BB, h}$ & $0.29_{-0.24}^{+0.30}$ km & & & \\[0.1cm]  \hline
\multirow{4}{1.25in}{\centering \textsc{wabs*tbabs*(nsmaxg)}\\ (H atmosphere, $B=10^{10}G$)} & $f$ &	$0.20 \pm 0.03$ &	\multirow{4}{0.8in}[-0.3cm]{\centering $30.46$} & \multirow{4}{0.8in}[-0.3cm]{\centering $56.06$} &	\multirow{4}{0.8in}[0.0cm]{\centering $30.43/16$ \\ $(0.016)$}\\[0.1cm]
	& $N_{H, SMC}$ & $2_{-2}^{+4*} \times 10^{21}$ cm$^{-2}$ & & & \\[0.1cm]
    & $T_{eff}$ & $1.7_{-0.1}^{0.4} \times 10^{6}$ K & & & \\[0.1cm]
    & $(R_{em}/R_{NS})$ & $1.0_{-0.5}^{+0.0**}$ & & & \\[0.1cm] \hline
\multirow{6}{1.25in}{\centering \textsc{wabs*tbabs*(nsmaxg + pegpwrlw)}\\ (H atmosphere, $B=10^{10}G$)} & $f$ &	$0.20 \pm 0.03$ &	\multirow{6}{0.8in}[-0.3cm]{\centering $25.40$} & \multirow{6}{0.8in}[-0.3cm]{\centering $56.25$} &	\multirow{6}{0.8in}[-0.2cm]{\centering $26.25/15$\\($0.036$)} \\[0.1cm]
	& $N_{H, SMC}$ & $1_{-1}^{+4*} \times 10^{21}$ cm$^{-2}$ & & & \\[0.1cm]
    & $T_{eff}$ & $1.6_{-0.1}^{+0.2} \times 10^{6}$ K & & & \\[0.1cm]
    & $(R_{em}/R_{NS})$ & $1.0_{-0.3}^{+0.0**}$ & & & \\[0.1cm]
	& $\Gamma$ &	2.0 & & & \\[0.1cm]
    & Flux$_{pl}$ & $(1.6_{-1.2}^{+1.4}) \times 10^{-15}$ ergs cm$^{-2}$ s$^{-1}$ & & & \\[0.1cm] \hline
\multirow{4}{1.25in}{\centering \textsc{wabs*tbabs*(nsmaxg)}\\ (H atmosphere, $B=10^{12}G$)} & $f$ &	$0.20 \pm 0.03$ & \multirow{4}{0.8in}[0.0cm]{\centering $31.64$} & \multirow{4}{0.8in}[-0.0cm]{\centering $57.24$} &	\multirow{4}{0.8in}[0.0cm]{\centering $31.67/16$\\($0.011$)}\\[0.1cm]
	& $N_{H, SMC}$ & $2_{-2}^{+4*} \times 10^{21}$ cm$^{-2}$ & & & \\[0.1cm]
    & $T_{eff}$ & $1.74_{-0.06}^{+0.42} \times 10^{6}$ K & & & \\[0.1cm]
    & $(R_{em}/R_{NS})$ & $1.0_{-0.5}^{+0.0**}$ & & & \\[0.1cm] \hline
\multirow{6}{1.25in}{\centering \textsc{wabs*tbabs*(nsmaxg + pegpwrlw)}\\ (H atmosphere, $B=10^{12}G$)} & $f$ &	$0.20_{-0.04}^{0.03}$ & \multirow{6}{0.8in}[0.0cm]{\centering $26.86$} & \multirow{6}{0.8in}[-0.0cm]{\centering $57.72$} &		\multirow{6}{0.8in}[0.0cm]{\centering $27.87/15$\\($0.022$)}\\[0.1cm]
	& $N_{H, SMC}$ & $1_{-1}^{+4*} \times 10^{21}$ cm$^{-2}$ & & & \\[0.1cm]
    & $T_{eff}$ & $1.7_{-0.1}^{+0.2} \times 10^{6}$ K & & & \\[0.1cm]
    & $(R_{em}/R_{NS})$ & $1.0_{-0.3}^{+0.0**}$ & & & \\[0.1cm]
	& $\Gamma$ &	2.0 & & & \\[0.1cm]
    & Flux$_{pl}$ & $(1.6 \pm 1.3) \times 10^{-15}$ ergs cm$^{-2}$ s$^{-1}$ & \\[0.1cm] \hline
\end{tabular}\\
\justify
Note: All the above models were added to the previously specified background model and fit to the observed X-ray spectra without background subtraction. ``f" represents the normalisation constant multiplied to the background model before being added to the above models. ``Flux$_{pl}$" represents the unabsorbed flux from the power-law component between 0.2 and 10.0 keV. We fixed the absorption column due to Milky way in the direction of SMC to $5.36 \times 10^{20}$ cm$^{-2}$ \citep{plucinsky2017}. We see that the source is best modelled by a neutron star with a carbon atmosphere, or a simple black body with a harder power-law or a hotter black body component. However, due to the high contribution from the background, we cannot conclusively reject other models used in our analysis. \\
$^{*}$ Lower error bound reaches the lower hard limit. \newline
$^{**}$ Upper error bound exceeds the upper hard limit. \newline
\end{table*}

\begin{table*}
    \centering
    \contcaption{Summary of modelling the combined spectrum of the compact object in SNR 1E 0102.2-7219.}
    \label{table:spec_e0102_cont}
    \begin{tabular}{c  c  c  c  c   c}
        \hline
        \multirow{2}{1.25in}[0.0cm]{\bf Model}  &	\multirow{2}{*}[0.0cm]{\bf Parameters} & \multirow{2}{*}[0cm]{\bf Parameter Values} &	\multirow{2}{0.8in}[0.0cm]{ \bf C-statistic} & \multirow{2}{0.8in}[0.0cm]{\bf AICc} & {\bf $\mathbf{\chi^2/d.o.f}$}  \\
        & & & & & {\bf (p-value)}\\
        \hline
        \multirow{4}{1.25in}{\centering \textsc{wabs*tbabs*(nsmaxg)}\\ (H atmosphere, $B=10^{13}G$)} & $f$ &	$0.20_{-0.04}^{+0.03}$ & \multirow{4}{0.8in}[0.0cm]{\centering $28.99$} & \multirow{4}{0.8in}[-0.0cm]{\centering $54.49$} &		\multirow{4}{0.8in}[0.0cm]{\centering $29.47/16$\\($0.021$)} \\[0.1cm]
	& $N_{H, SMC}$ & $2_{-2}^{+4*} \times 10^{21}$ cm$^{-2}$ & & & \\[0.1cm]
    & $T_{eff}$ & $1.8_{-0.1}^{+0.3} \times 10^{6}$ K & & & \\[0.1cm]
    & $(R_{em}/R_{NS})^2$ & $1.0_{-0.4}^{+0.0**}$ & & & \\[0.1cm] \hline
        \multirow{6}{1.25in}{\centering \textsc{wabs*tbabs*(nsmaxg + pegpwrlw)}\\ (H atmosphere, $B=10^{13}G$)} & $f$ &	$0.19 \pm 0.03$ & \multirow{6}{0.8in}[0.0cm]{\centering $25.52$} & \multirow{6}{0.8in}[-0.0cm]{\centering $56.37$} &	\multirow{6}{0.8in}[0.0cm]{\centering $26.37/15$\\($0.034$)} \\[0.1cm]
	& $N_{H, SMC}$ & $2_{-2}^{+3*} \times 10^{21}$ cm$^{-2}$ & & &\\[0.1cm]
    & $T_{eff}$ & $1.74_{-0.08}^{+0.19} \times 10^{6}$ K & & &\\[0.1cm]
    & $(R_{em}/R_{NS})$ & $1.0_{-0.3}^{+0.0**}$ & & &\\[0.1cm]
	& $\Gamma$ &	2.0 & & &\\[0.1cm]
    & Flux$_{pl}$ & $(1.4_{-1.3}^{+1.4}) \times 10^{-15}$ ergs cm$^{-2}$ s$^{-1}$ & & & \\[0.1cm] \hline
\multirow{4}{1.25in}{\centering \textsc{wabs*tbabs*(nsmaxg)}\\ (C atmosphere, $B=10^{12}G$)} & $f$ &	$0.21_{-0.04}^{+0.03}$ & \multirow{4}{0.8in}[0.0cm]{\centering $19.08$} & \multirow{4}{0.8in}[-0.0cm]{\centering $44.69$} &	\multirow{4}{0.8in}[0.0cm]{\centering $18.85/16$\\($0.28$)} \\[0.1cm]
	& $N_{H, SMC}$ & $9_{-7}^{+12} \times 10^{21}$ cm$^{-2}$ & & &\\[0.1cm]
    & $T_{eff}$ & $3.0_{-0.4}^{+0.5} \times 10^{6}$ K & & &\\[0.1cm]
    & $(R_{em}/R_{NS})$ & $0.5_{-0.2}^{+0.5}$ & & &\\[0.1cm] \hline
    \end{tabular}
\end{table*}

We also check the validity of our method by applying this analysis technique for individual background regions. We extract a spectrum from each background region, and see if each spectrum can be fit by a simple scaling of the combined background model discussed in Table~\ref{tab:bkg_spec}. We find that the $\chi^2$/d.o.f for background regions 1-6 are $15.24/13$, $6.38/12$, $22.67/10$, $21.81/16$, $9.87/13$ \& $22.41/11$, respectively. These fits are much better than our fit of the source spectrum with the combined background model ($p \gtrsim 0.15$, except for regions 3 and 6 which have $p = 0.01, 0.02$, respectively.)

We next analyse the source spectrum by adding different models to our background model. We maintain the two-component absorption model, with fixed Galactic absorption (\texttt{wabs})(to $5.36 \times 10^{20}$ cm$^{-2}$), but allow the SMC (\texttt{tbabs}) absorption to vary. The results of our analysis are summarised in Table~\ref{table:spec_e0102}. The simple power-law gives $N_{H, SMC} = 9_{-4}^{+8} \times 10^{21}$ cm$^{-2}$ and $\Gamma = 4.6_{-0.5}^{+0.9}$, with a C-statistic value of 22.66. Though this model is a better fit to the data with $\chi^{2}/$d.o.f. $= 23.93/16$ ($p=0.091$), such a large value of $\Gamma$ has not been seen for non-thermal emission from pulsars or pulsar wind nebulae \citep{Li08}, but is typical of the values found when power-law models are fit to spectra better described by low-temperature blackbody (BB) or BB-like spectra, typical of NSs.

\begin{figure*}
    \centering
    \begin{subfigure}[b]{0.45\textwidth}
        \includegraphics[width=\textwidth]{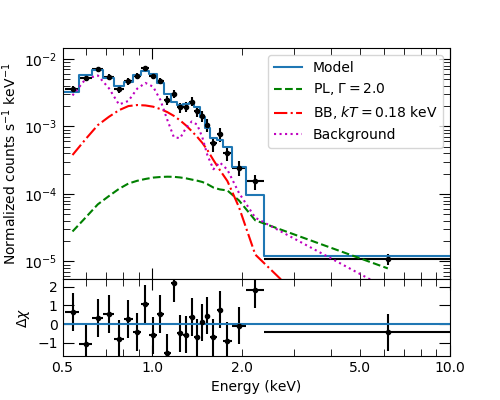}
        \caption{Blackbody + power-law fit to the observed spectrum. The source and  background are modelled simultaneously. The best fitting model gives $N_{H, SMC} = 5.6_{-5.1}^{+10.7} \times 10^{21}$~cm$^{-2}$, $T_{BB} = 2.1_{-0.3}^{+0.4} \times 10^6$~K, $R_{BB} = 8_{-4}^{+7}$~km. The power-law photon index is fixed to 2.0. The model gives $\chi^2/\mathrm{d.o.f} = 17.33/15$, and $p = 0.30$, indicating a good fit.}
        \label{fig:bbody_pegpw}
    \end{subfigure}
    \hfill
    \begin{subfigure}[b]{0.45\textwidth}
        \centering
        \includegraphics[width=\textwidth]{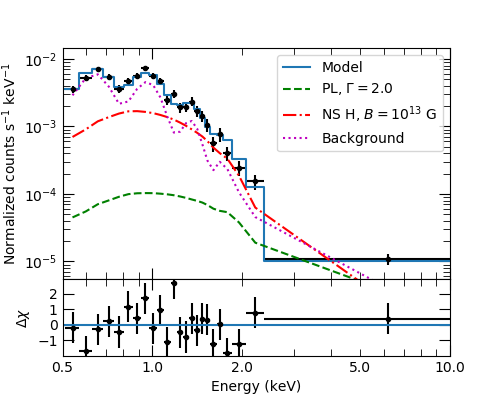}
        \caption{NS H atmosphere + power-law fit to the source spectrum and background. We use $M_{NS} = 1.4$~M$_{\odot}$, $R_{NS} = 12$~km, $B = 10^{13}$~G, and photon index $\Gamma = 2.0$ for this  spectral fit. The best fit has $N_{H, SMC} = 2_{-2}^{+3} \times  10^{21}$~cm$^{-2}$, $T_{eff} = 1.74_{-0.08}^{+0.19}\times10^6$~K, and $R_{em}/R_{NS} = 1.0_{-0.3}^{+0.0}$. With $\chi^2/\mathrm{d.o.f} = 25.52/15$, giving $p = 0.034$, this is a poor fit.}
        \label{fig:ns13_pw}
    \end{subfigure}

    \begin{subfigure}[b]{0.45\textwidth}
        \centering
        \includegraphics[width=\textwidth]{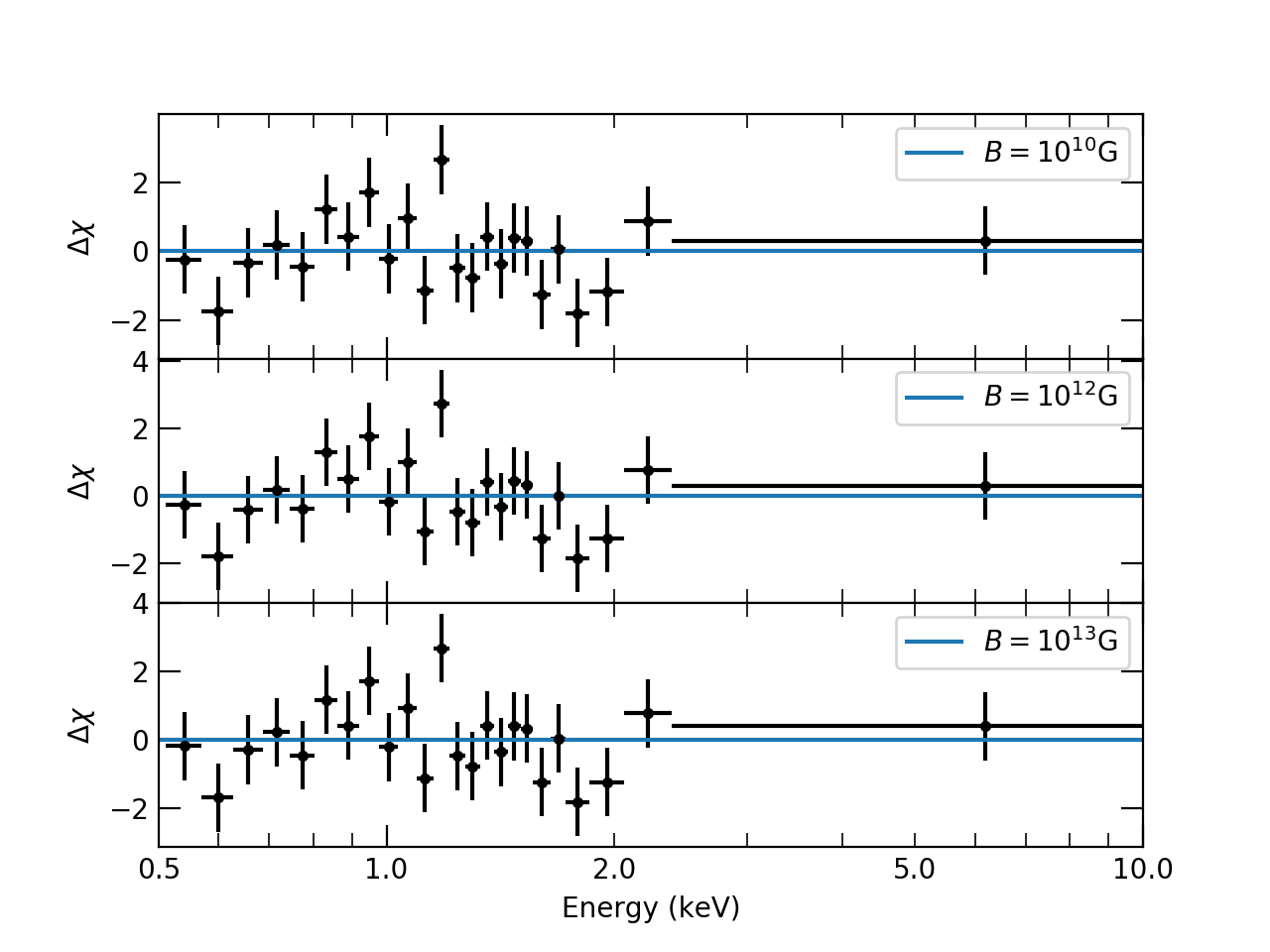}
        \caption{Residuals to fits with a thermal NS H atmosphere and non-thermal power-law, for three  different surface magnetic fields ($B=10^{10},10^{12},10^{13}$~G). The three models have similar residuals (at $\sim$1 and $\sim$1.8~keV), AICc values ($\sim 57$), and $\chi^2$ ($\sim 27$ for 15 d.o.f, i.e. p-value $\sim 0.02 - 0.03)$. }
        \label{fig:nsH_residuals}
    \end{subfigure}
    \hfill
    \begin{subfigure}[b]{0.45\textwidth}
        \centering
        \includegraphics[width=\textwidth]{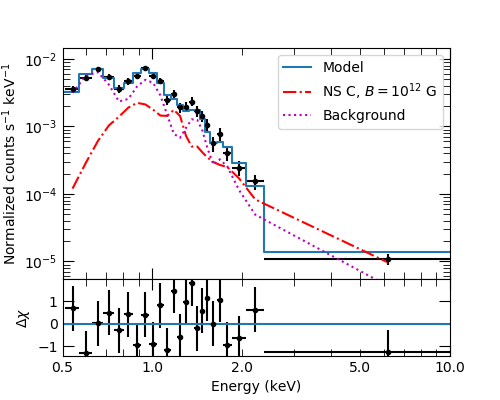}
        \caption{NS C atmosphere fit to the source spectrum on top of the background model. We use  $M_{NS} = 1.4$~M$_{\odot}$, $R_{NS} = 12$~km, and $B = 10^{12}$~G. The best fitting model has $N_{H, SMC} = 9_{-7}^{+12} \times10^{21}$~cm$^{-2}$, $T_{eff} = 3.0_{-0.4}^{+0.5}\times10^6$~K, and $R_{em}/R_{NS} = 0.5_{-0.2}^{+0.5}$. This model does not need a power-law component to explain emission at high energies. With $\chi^2/\mathrm{d.o.f} = 19.08/16$, giving $p = 0.22$, this is a good fit.}
        \label{fig:ns12C}
    \end{subfigure}
    \caption{Spectral analysis of thermal component in E0102 using (a) BB+PL, (b) and (c) NS H atmosphere and (d) NS C atmosphere models. BB+PL and NS C atmosphere at $B=10^{12}$ G best fit the observed spectra, while NS H atmosphere models give a poor fit.}
\end{figure*}

We therefore consider BB and BB-like models for the spectral fit. We first fit with the \texttt{bbodyrad} model in XSPEC. This gives $N_{H, SMC} = 2_{-2}^{+8} \times 10^{21}$ cm$^{-2}$, $T_{BB} = (2.9 \pm 0.6) \times 10^6$ K, $R_{BB} = 3_{-1}^{+4}$ km and C-statistic value of $34.89$. This model has $\chi^2/$d.o.f. $= 32.41/16$ ($p = 0.0088$), a poor fit, and fails to explain the emission at higher energies. 

Many pulsars show comparable non-thermal X-ray emission, from a magnetosphere and/or pulsar wind nebula (PWN), as thermal X-ray emission from the NS surface \citep[e.g.][]{pavlov2004a}. Therefore, we add a power-law with $\Gamma$ fixed to $2$ \citep[typical of pulsar wind nebulae,][]{Li08} to include possible non-thermal emission from the magnetosphere and/or PWN. (We also tried allowing the power-law index to vary, but this led to the fitting parameters being poorly constrained, without improving the final fit.)  As the black body fit is consistent with X-ray emission from a NS, we constrain the maximum radius of the emitting region to 15 km. 
The best fitting model, shown in Fig.~\ref{fig:bbody_pegpw}, has $N_{H, SMC} = 5.6_{-5.1}^{+10.7} \times 10^{21}$ cm$^{-2}$, $T_{BB} = 2.1_{-0.8}^{+0.4} \times 10^6$ K, $R_{BB} = 8_{-4}^{+7}$ km. The flux from the power-law component, Flux$_{pl} = (3 \pm 1) \times 10^{-15}$ ergs cm$^{-2}$ s$^{-1}$ between 0.5 and 10 keV, and the C-statistic value is $18.11$. This model gives $\chi^2/$d.o.f $= 17.33/15$ i.e. $p = 0.30$, indicating an adequate fit. The unabsorbed 0.5-10 keV thermal luminosity, $L_{BB} = 6_{-2}^{+11} \times 10^{33}$ ergs s$^{-1}$, is greater than the non-thermal luminosity, $L_{pl} = (1.4 \pm 0.4) \times 10^{33}$ ergs s$^{-1}$.

The harder X-ray component could also be fit with a hotter blackbody of temperature, $T_{BB, h} = 6_{-1}^{+8} \times 10^6$ K and radius, $R_{BB, h} = 0.05 - 0.6$ km. This model could be interpreted as indicating the presence of hotspots at the magnetic poles of the NS. This model has similar C- \& $\chi^2$ statistics as the BB+PL model, but a higher AICc statistic due to one additional fitting parameter. We note that the best fit temperature is at the high end of the observed temperatures on NS (see Sec~\ref{sec:disc}). Similar to the previous case, the softer component (with luminosity, $L_{x, c} = 1.3_{-0.9}^{+0.8} \times 10^{34}$ ergs s$^{-1}$) tends to be more luminous than the harder component (with luminosity, $L_{x, h} = 1.0_{-0.5}^{+0.6} \times 10^{33 }$ ergs s$^{-1}$).

To constrain the range of possible natures of this NS, we then try fits with a variety of NS atmosphere models.
We use the \texttt{nsmaxg} model in XSPEC which has spectral templates for different magnetic fields for hydrogen atmospheres, as well as for heavier elements  \citep{mori2007, ho2008}. We fix the NS mass to 1.4 M$_{\odot}$ and radius to 12 km and vary the temperature and the normalisation of the NS to fit the spectra. We fix the upper limit of the normalisation parameter to 1, indicating emission from the entire surface. We tried models with magnetic fields $B = 10^{10}$, $10^{12}$, and $10^{13}$ G. These models fail to explain the emission at hard X-rays ($E > 2$ keV; $p \sim 0.01 -0.02$ for all NS H atmosphere models). Therefore we add a power-law component with fixed photon index, 2, to the spectra to model possible magnetospheric or PWN emission. Adding a power-law reduces the C-statistic by $\sim 4$ while decreasing one degree of freedom (i.e. no significant change in AICc statistic, as seen in Table~\ref{table:spec_e0102}). With the power-law, the best fitting models with different $B$ have similar C-statistic values ($\sim 25 - 27$ for 15 d.o.f), $N_{H, SMC}$ ($\sim 2 \times  10^{21}$ cm$^{-2}$), effective temperatures, ($T_{eff}$ $\sim 1.7 \times 10^{6}$ K) and radius of the emitting region ($R_{em}/R_{NS}$ $\sim 1$). These models give a thermal luminosity $L_{X, th} \sim 4 \times 10^{33}$ ergs s$^{-1}$ in the $0.5-10.0$ keV range.  The large $\chi^2 \sim 26-27$ values for 15 d.o.f (i.e $p \sim 0.02 - 0.03$) indicate that NS H atmosphere models do not fit the data well, even after the addition of a non-thermal power-law component. We show the spectral fit with $B =  10^{13}$ G in Fig.~\ref{fig:ns13_pw}, and residuals to the spectral fits for all three models in Fig.~\ref{fig:nsH_residuals}.

Note that a NS H atmosphere spectrum with a given effective temperature, $T_{eff}$, is much harder than that of a BB with the same BB temperature $T_{BB}$, since the emission at higher energies is from hotter layers of the hydrogen atmosphere \citep[e.g.][]{Zavlin96}. Thus, while the BB fit has a higher temperature and might indicate the presence of hotpots, the NS hydrogen atmosphere fits have a lower temperature and suggest emission from the complete NS surface. All the hydrogen atmosphere models leave significant residuals around 1 keV. As the hydrogen atmosphere models have broader spectra than the blackbody models (which are more sharply curved), this may be indicating a preference for a blackbody-like shape over the hydrogen atmosphere model shapes.
However, considering the complex and bright background, we should consider whether these residuals are caused by complexities in the background subtraction (see \S~\ref{sec:back}).

If these spectral residuals at $\sim 1$ keV are indeed due to a sharper peak in the data than that of the absorbed NS H atmosphere model, they cannot be explained using multiple NS H atmospheres. Allowing for a lower temperature, $T_{eff} \sim  8 \times 10^5$ K and larger absorption can replicate a narrower peak, but such a model would require an emitting region of $R_{em} > 25$ km, which is not feasible for a neutron star. Since heavier element atmospheres can have different spectral slopes  \citep[for example, a C atmosphere shows a sharper decline at $\sim 1-2$ keV as compared to the H atmospheres;][]{mori2007}, we try to fit the observed spectrum using the C, O and Ne templates provided in \texttt{nsmaxg}.

We find that a carbon atmosphere with $B = 10^{12}$ G fits the X-ray spectrum  best. This model can also explain the emission at the higher energies without an additional power-law component. Fig.~\ref{fig:ns12C} shows the best fitting C atmosphere model. This model gives $N_H = 9_{-7}^{+12} \times 10^{21}$ cm$^{-2}$, $T_{eff} = 3.0_{-0.4}^{+0.5} \times 10^6$ K and $R_{em}/R_{NS} = 0.5_{-0.2}^{+0.5}$ with a C-statistic value of $19.08$. With  $\chi^2/\mathrm{d.o.f} = 18.85/16$, i.e. $p = 0.28$ (which is comparable in quality to the \texttt{blackbody+pegpwrlw} fit), this model is an adequate fit. This model gives a thermal luminosity, $L_{X, th} = 1.1_{-0.5}^{+1.6} \times 10^{34}$ ergs s$^{-1}$ between $0.5$ and $10$ keV. This fit is $\sim 70$ times better than the fit with the $B=10^{13}$ G hydrogen atmosphere model, and $\sim 8$ times better than the fit to the  \texttt{blackbody+pegpwrlw} model, based on AICc statistics. Using a non-magnetic carbon atmosphere model \citep[e.g., \texttt{carbatm};][]{suleimanov14}, or a $B = 10^{13}$ G carbon atmosphere worsens the fit. Using heavier elements like Ne or O gives no improvement over the hydrogen atmosphere models. Thus the spectral analysis favours a NS with a $10^{12}$ G carbon atmosphere for the compact object in this SNR, though the complexities of background subtraction must be carefully considered (see \S~\ref{sec:back}).

\begin{table*}
\centering
\caption{Summary of spectral analysis by simultaneously modelling the the individual spectra of compact object in SN 1E 0102.2-7219.}
\label{table:spec_e0102_sim}
\begin{tabular}{c  c  c  c  c c}
\hline
\multirow{2}{*}[0.0cm]{\bf Model}  &	\multirow{2}{*}[0.0cm]{\bf Parameters} & \multirow{2}{*}[0cm]{\bf Parameter Values} &	\multirow{2}{0.8in}[0.0cm]{\centering \bf C-statistic\\(d.o.f)} & \multirow{2}{*}[0.0cm]{\bf AICc} & {\bf log (CvM)}  \\
& & & & & {\bf (goodness)}\\
\hline
\multirow{4}{1.3in}[0.0cm]{\centering \textsc{wabs*tbabs*bbodyrad}} & $f$ &	$0.20_{-0.02}^{+0.04}$ & \multirow{4}{0.8in}[0.0cm]{\centering $577.56$\\($675$)} & \multirow{4}{0.8in}[0.0cm]{\centering $593.77$} & \multirow{4}{0.8in}[0.0cm]{\centering $-10.80$ \\ ($3\%$)}\\[0.1cm]
    & $N_{H, SMC}$ & $< 3^{*} \times 10^{22}$ cm$^{-2}$ & & &\\[0.1cm]
    & $T_{BB}$ & $3.1_{-0.4}^{+0.6} \times 10^6$ K & & &\\[0.1cm]
    & $R_{BB}$ & $3 \pm 1$ km & & & \\[0.1cm] \hline
 \multirow{6}{1.3in}[-0.2cm]{\centering \textsc{wabs*tbabs*(bbodyrad + pegpwrlw)}} & $f$ & $0.16_{-0.04}^{+0.05}$ & \multirow{6}{0.8in}[0.0cm]{\centering $557.27$ \\ ($674$)} & \multirow{6}{0.8in}[0.0cm]{\centering $575.54$} & \multirow{6}{0.8in}[0.0cm]{\centering $-10.85$ \\ ($2\%$)} \\[0.1cm]
    & $N_{H, SMC}$ & $3_{-3}^{+11*} \times 10^{21}$ cm$^{-2}$ & & & \\[0.1cm]
    & $T_{BB}$ & $2.1_{-0.4}^{+0.5} \times 10^6$ K & & & \\[0.1cm]
    & $R_{BB}$ & $7_{-3}^{+8**}$ km & & & \\[0.1cm]
	& $\Gamma$ &	$2.0$ & & & \\[0.1cm]
    & Flux$_{pl}$ & $ (5 \pm 2) \times 10^{-15}$ ergs cm$^{-2}$ s$^{-1}$ & & & \\[0.1cm] \hline
\multirow{6}{1.3in}[-0.2cm]{\centering \textsc{wabs*tbabs*(bbodyrad + bbodyrad)}} & $f$ & $0.16_{-0.04}^{+0.05}$ & \multirow{6}{0.8in}[0.0cm]{\centering $560.97$ \\ ($673$)} & \multirow{6}{0.8in}[0.0cm]{\centering $581.29$} & \multirow{6}{0.8in}[0.0cm]{\centering $-10.83$ \\ ($2\%$)} \\[0.1cm]
    & $N_{H, SMC}$ & $2_{-2}^{+10*} \times 10^{21}$ cm$^{-2}$ & & & \\[0.1cm]
    & $T_{BB, c}$ & $ 2.2_{-0.5}^{+0.6} \times 10^6$ K & & & \\[0.1cm]
    & $R_{BB, c}$ & $6_{-1}^{+5}$ km & & & \\[0.1cm]
	& $T_{BB, h}$ & $ > 5 \times 10^6$ K & & & \\[0.1cm]
    & $R_{BB, h}$ & $< 0.29$ km & & & \\[0.1cm] \hline
\multirow{4}{1.3in}{\centering \textsc{wabs*tbabs*(nsmaxg)}\\ (H atmosphere, $B=10^{10}G$)} & $f$ &	$0.19_{-0.03}^{+0.04}$ &	\multirow{4}{0.8in}[-0.3cm]{\centering $573.32$\\($675$)} & \multirow{4}{0.8in}[-0.3cm]{\centering $589.53$} &	\multirow{4}{0.8in}[0.0cm]{\centering $-10.82$ \\ $(2\%)$}\\[0.1cm]
	& $N_{H, SMC}$ & $ < 5^{*} \times 10^{21}$ cm$^{-2}$ & & & \\[0.1cm]
    & $T_{eff}$ & $1.8_{-0.2}^{+0.5} \times 10^{6}$ K & & & \\[0.1cm]
    & $(R_{em}/R_{NS})$ & $0.9_{-0.3}^{+0.1**}$ & & & \\[0.1cm] \hline
\multirow{6}{1.3in}{\centering \textsc{wabs*tbabs*(nsmaxg + pegpwrlw)}\\ (H atmosphere, $B=10^{10}G$)} & $f$ &	$0.18_{-0.02}^{+0.04}$ &	\multirow{6}{0.8in}[-0.3cm]{\centering $562.42
$\\($674$)} & \multirow{6}{0.8in}[-0.3cm]{\centering $580.69$} &	\multirow{6}{0.8in}[-0.2cm]{\centering $-10.78$\\($3\%$)} \\[0.1cm]
	& $N_{H, SMC}$ & $< 3^{*} \times 10^{21}$ cm$^{-2}$ & & & \\[0.1cm]
    & $T_{eff}$ & $1.6_{-0.1}^{+0.3} \times 10^{6}$ K & & & \\[0.1cm]
    & $(R_{em}/R_{NS})$ & $1.0_{-0.3}^{+0.0**}$ & & & \\[0.1cm]
	& $\Gamma$ &	2.0 & & & \\[0.1cm]
    & Flux$_{pl}$ & $(3 \pm 2) \times 10^{-15}$ ergs cm$^{-2}$ s$^{-1}$ & & & \\[0.1cm] \hline
\multirow{4}{1.3in}{\centering \textsc{wabs*tbabs*(nsmaxg)}\\ (H atmosphere, $B=10^{12}G$)} & $f$ &	$0.19_{-0.03}^{+0.04}$ & \multirow{4}{0.8in}[0.0cm]{\centering $574.35$\\($675$)} & \multirow{4}{0.8in}[-0.0cm]{\centering $590.56$} &	\multirow{4}{0.8in}[0.0cm]{\centering $-10.82$\\($2\%$)}\\[0.1cm]
	& $N_{H, SMC}$ & $2_{-2}^{+4*}) \times 10^{21}$ cm$^{-2}$ & & & \\[0.1cm]
    & $T_{eff}$ & $1.8_{-0.1}^{+0.6} \times 10^{6}$ K & & & \\[0.1cm]
    & $(R_{em}/R_{NS})$ & $1.0_{-0.6}^{+0.0**}$ & & & \\[0.1cm] \hline
\multirow{6}{1.3in}{\centering \textsc{wabs*tbabs*(nsmaxg + pegpwrlw)}\\ (H atmosphere, $B=10^{12}G$)} & $f$ &	$0.18_{-0.03}^{+0.04}$ & \multirow{6}{0.8in}[0.0cm]{\centering $563.66$\\($674$)} & \multirow{6}{0.8in}[-0.0cm]{\centering $581.92$} &		\multirow{6}{0.8in}[0.0cm]{\centering $-10.75$\\($3\%$)}\\[0.1cm]
	& $N_{H, SMC}$ & $(1_{-1}^{+3*}) \times 10^{21}$ cm$^{-2}$ & & & \\[0.1cm]
    & $T_{eff}$ & $1.7_{-0.1}^{+0.3} \times 10^{6}$ K & & & \\[0.1cm]
    & $(R_{em}/R_{NS})$ & $1.0_{-0.3}^{+0.0**}$ & & & \\[0.1cm]
	& $\Gamma$ &	2.0 & & & \\[0.1cm]
    & Flux$_{pl}$ & $(3 \pm 2) \times 10^{-15}$ ergs cm$^{-2}$ s$^{-1}$ & \\[0.1cm] \hline
\end{tabular}
\justify
Note: Variables and notations used have the same meaning as in Table~\ref{table:spec_e0102}. The last column indicates the value of the Cramer von-Mises statistic \citep{cramer1928, vonMises1928} and the goodness value corresponding to it. The goodness value indicates the fraction of the realisations of the model which have a CvM statistic smaller than that of the data. A large value (say 95\%) indicates that the observed spectrum can be rejected (with 95\% confidence). We note that the best fitting values of the parameters are similar to their corresponding values in Table~\ref{table:spec_e0102}.
\end{table*}

\begin{table*}
    \centering
    \contcaption{Summary of spectral analysis by simultaneously modelling the the individual spectra of compact object in SN 1E 0102.2-7219.}
    \label{table:spec_e0102_sim_cont}
    \begin{tabular}{c  c  c  c  c   c}
        \hline
    	\multirow{2}{*}[0.0cm]{\bf Model}  &	\multirow{2}{*}[0.0cm]{\bf Parameters} & \multirow{2}{*}[0cm]{\bf Parameter Values} &	\multirow{2}{0.8in}[0.0cm]{\centering \bf C-statistic\\(d.o.f)} & \multirow{2}{*}[0.0cm]{\bf AICc} & {\bf log (CvM)}  \\
        & & & & & {\bf (goodness)}\\
        \hline
        \multirow{4}{1.3in}{\centering \textsc{wabs*tbabs*(nsmaxg)}\\ (H atmosphere, $B=10^{13}G$)} & $f$ &	$0.19 \pm 0.04$ & \multirow{4}{0.8in}[0.0cm]{\centering $571.99$\\($675$)} & \multirow{4}{0.8in}[-0.0cm]{\centering $588.20$} &		\multirow{4}{0.8in}[0.0cm]{\centering $-10.82/16$\\($2\%$)} \\[0.1cm]
	& $N_{H, SMC}$ & $1_{-1}^{+5*} \times 10^{21}$ cm$^{-2}$ & & & \\[0.1cm]
    & $T_{eff}$ & $1.9_{-0.1}^{+0.4} \times 10^{6}$ K & & & \\[0.1cm]
    & $(R_{em}/R_{NS})^2$ & $0.9_{-0.4}^{+0.1**}$ & & & \\[0.1cm] \hline
 \multirow{6}{1.3in}{\centering \textsc{wabs*tbabs*(nsmaxg + pegpwrlw)}\\ (H atmosphere, $B=10^{13}G$)} & $f$ &	$0.18_{-0.03}^{0.04}$ & \multirow{6}{0.8in}[0.0cm]{\centering $561.78$\\($674$)} & \multirow{6}{0.8in}[-0.0cm]{\centering $580.04$} &	\multirow{6}{0.8in}[0.0cm]{\centering $-10.79$\\($3\%$)} \\[0.1cm]
	& $N_{H, SMC}$ & $(4_{-4}^{+15}) \times 10^{21}$ cm$^{-2}$ & & &\\[0.1cm]
    & $T_{eff}$ & $1.7_{-0.1}^{+0.3} \times 10^{6}$ K & & &\\[0.1cm]
    & $(R_{em}/R_{NS})$ & $1.0_{-0.3}^{+0.0**}$ & & &\\[0.1cm]
	& $\Gamma$ &	2.0 & & &\\[0.1cm]
    & Flux$_{pl}$ & $(3 \pm 2) \times 10^{-15}$ ergs cm$^{-2}$ s$^{-1}$ & & & \\[0.1cm] \hline
\multirow{4}{1.3in}{\centering \textsc{wabs*tbabs*(nsmaxg)}\\ (C atmosphere, $B=10^{12}G$)} & $f$ &	$0.18_{-0.04}^{+0.05}$ & \multirow{4}{0.8in}[0.0cm]{\centering $560.44$\\($675$)} & \multirow{4}{0.8in}[-0.0cm]{\centering $576.65$} &	\multirow{4}{0.8in}[0.0cm]{\centering $-10.82$/16\\($2\%$)} \\[0.1cm]
	& $N_{H, SMC}$ & $3_{-3}^{+17} \times 10^{21}$ cm$^{-2}$ & & &\\[0.1cm]
    & $T_{eff}$ & $3.4_{-0.7}^{+0.3} \times 10^{6}$ K & & &\\[0.1cm]
    & $(R_{em}/R_{NS})$ & $0.3_{-0.1}^{+0.4}$ & & &\\[0.1cm] \hline
    \end{tabular}
\end{table*}

\subsection{Simultaneous spectral analysis for individual years}
\label{sec:sim_an}

We also check if the changing response of the {\it Chandra} ACIS instrument over time affects the results of our  spectral analyses significantly. As the ACIS instrument changes slowly over time, we sum spectra within each calendar year to retain sufficient statistics. We load the spectra from individual years separately into XSPEC, and fit them simultaneously using the models discussed above. We summarise the results of our simultaneous spectral analysis in Table~\ref{table:spec_e0102_sim}.

We note that both methods of spectral analysis (fitting the combined spectrum, and simultaneously fitting spectra loaded separately) result in similar parameter values for the different models used. The AICc statistics for the models used follow a similar trend,  with the \texttt{(BB+PL)}, and \texttt{NSMAXG} with $B=10^{12}G$ C atmosphere, giving the best fits (both these models have similar AICc values). We also note that the change in the C-statistics by the addition of a PL or hotter BB to the single BB model is roughly equal in both cases ($\sim 17$). The best fitting C atmosphere model is $\sim 6$ times better than the $10^{13}$ G H atmosphere model with a  power-law (and $\sim 200$ times better than the H atmosphere NS model alone). Thus, a blackbody plus non-thermal power-law, or a neutron star with a carbon atmosphere, are the most favourable fits.  However, this method of spectral analysis doesn't give much insight into the quality of the individual fits, as XSPEC's goodness simulations indicate that $< 5\%$ of the realisations have CvM statistics smaller than the best fitting model in all cases. Another difference between the two methods of fitting is that adding a power-law to the H atmosphere models doesn't change the C-statistics and the AICc value appreciably ($\Delta cstat \sim 5$, $\Delta$AICc $< 1$ corresponding to $< 1.5$ times better) while analysing the combined spectrum, but when simultaneously modelling the individual spectra, $\Delta$ AICc $\sim 8$; i.e. adding a power-law improves the model by a factor of $\sim 50$.

\subsection{Effects of altering background selection} \label{sec:back}
Given the bright background, it is important to study how our choice of background spectrum affects our  spectral modelling. To study the effect of changes in the background on the spectral parameters, we first model the spectra of each individual background region shown in Fig. ~\ref{fig:regions}.  We use the background model described above (i.e., \texttt{wabs*tbabs*(apec + gaussian + gaussian + gaussian + gaussian)}) to model the individual background spectra. We find that grouping the background spectra such that each bin consists of at least one photon per bin, and fitting them using C-statistics, gives the best constraints on the parameter values. We then use these various background models to analyse the combined source spectrum (modelling the  background and source simultaneously). Using these different background models, we analyse the change in the best fitting parameters when the source is modelled using --- \texttt{wabs*tbabs*bbodyrad}, \texttt{wabs*tbabs*(bbodyrad + pegpwrlw)}, \texttt{wabs*tbabs*(nsmaxg + pegpwrlw)} (H atmosphere, B=$10^{13}G$) and  \texttt{wabs*tbabs*nsmaxg} (C atmosphere, B=$10^{12}G$). Our results are summarised in Table~\ref{table:bkg_effect}. 

In general, we notice that all the best fitting parameters stay within the same error limits even when the underlying background is modelled differently. However, we do notice that the C-statistics and $\chi^2$ of the best fitting models change significantly.  Inspecting the source spectra where \texttt{BB + PL} and \texttt{nsmaxg} with $B=10^{12}$G C atmosphere models are not good fits, shows that varying the normalisation and width of the background emission lines near the residuals significantly changes the fit quality. 
For example, when the background is modelled from region 1 alone, fixing $\sigma_{Mg}$ to 0.07 (from Table~\ref{tab:bkg_spec}) and allowing Flux$_{Mg}$ to vary reduces the C-statistic by 12.72, while decreasing d.o.f by 1 (i.e $\sim 50$ times better according to AICc),  giving $\chi^2/d.o.f = 19.02/15$ (p-value = $0.21$, i.e. a good fit). This signifies that properly modelling the background is crucial to understand the quality of a spectral fit. However we do notice that in all cases, a blackbody model with a power-law and a neutron star with a C atmosphere are still better fits than a simple BB or a NS with a H atmosphere. The NS H atmosphere plus power-law model has a reduced $\chi^2$ ($\chi^2_{\nu}$) $ > 1.5$ for all the different background models used, due to residuals at $\sim 1$ keV. Thus these residuals seem to be real, and not due to incorrect modelling of the background.

\afterpage{
\clearpage
\begin{landscape}
\begin{table}
\centering
\caption{Best fitting spectral parameters for different background regions}
\label{table:bkg_effect}
\begin{tabular}{c c c c c c c c}
    \hline
    \multirow{2}{*}{\bf Model} & \multirow{2}{*}{\bf Parameters} & \multicolumn{6}{c}{\bf Parameter values on modelling background from} \\
    & & {\bf Region 1} & {\bf Region 2} & {\bf Region 3} & {\bf Region 4} & {\bf Region 5} & {\bf Region 6} \\[0.1cm]
    \hline
    \multirow{4}{1.5in}[0.0cm]{\centering \textsc{wabs*tbabs*bbodyrad}} & $f$ & $1.6 \pm 0.3$ &	$1.4 \pm 0.2$ & $1.5 \pm 0.2$ & $0.8 \pm 0.1$  & $1.3 \pm 0.2$ & $1.1_{-0.1}^{+0.2}$ \\[0.1cm]
    & $N_{H, SMC}$ & $3_{-3}^{+22*} \times 10^{21}$ cm$^{-2}$ & $2_{-2}^{+6*} \times 10^{21}$ cm$^{-2}$ & $ 1_{-1}^{+4*} \times 10^{21}$ cm$^{-2}$ & $ < 8^{*} \times 10^{21}$ cm$^{-2}$ & $1_{-1}^{+6*} \times 10^{21}$ cm$^{-2}$ & $< 4^{*} \times 10^{21}$ cm$^{-2}$ \\[0.1cm]
    & $T_{BB}$ & $2.5_{-0.9}^{+0.5} \times 10^6$ K & $(3.1 \pm 0.6) \times 10^6$ K & $3.1_{-0.3}^{+0.4} \times 10^6$ K & $2.8_{-0.5}^{+0.7} \times 10^6$ K & $(2.6 \pm 0.4) \times 10^6$ K & $(3.3 \pm 0.4) \times 10^6$ K \\[0.1cm]
    & $R_{BB}$ & $4_{-2}^{+7}$ km & $2_{-1}^{+3}$ km & $3_{-1}^{+2}$ km & $3_{-1}^{+3}$ km  & $4_{-1}^{+4}$ km & $2.4_{-0.7}^{+1.1}$ km \\[0.1cm]
    & C-statistic & $30.05$ & $37.47$ & $57.28$ & $33.29$ & $38.16$ & $44.45$ \\[0.1cm]
    & $\chi^2$/d.o.f & $30.18/16$ & $36.74/16$ & $56.92/16$ & $29.93/16$& $36.55/16$ & $44.29/16$ \\[0.1cm] \hline
    \multirow{4}{1.5in}[0.0cm]{\centering \textsc{wabs*tbabs*(bbodyrad + pegpwrlw}} & $f$ & $1.4_{-0.3}^{+0.4}$ & $1.3_{-0.3}^{+0.2}$ & $1.4_{-0.3}^{+0.2}$ & $0.6_{-0.1}^{+0.2}$ & $1.1_{-0.3}^{+0.2}$ & $1.0_{-0.3}^{+0.2}$ \\[0.1cm]
    & $N_{H, SMC}$ & $2_{-2}^{+13*} \times 10^{21}$ cm$^{-2}$ & $8_{-7}^{+8} \times 10^{21}$ cm$^{-2}$ & $1.0_{-0.6}^{+0.4} \times 10^{22}$ cm$^{-2}$ & $2_{-2}^{+7*} \times 10^{21}$ cm$^{-2}$ & $2_{-2}^{+6*} \times 10^{21}$ cm$^{-2}$ & $4.1_{-3.5}^{+8.6} \times 10^{21}$ cm$^{-2}$ \\[0.1cm]
    & $T_{BB}$ & $2.3_{-0.5}^{+0.3} \times 10^6$ K & $1.9_{-0.2}^{+0.6} \times 10^6$ K & $1.9_{-0.1}^{+0.3} \times 10^6$ K & $(2.2 \pm 0.4) \times 10^6$ K & $2.1_{-0.3}^{+0.4} \times 10^6$ K & $2.2_{-0.4}^{+0.3} \times 10^6$ K \\[0.1cm]
    & $R_{BB}$ & $6_{-3}^{+9}$ km & $11_{-7}^{+4**}$ km & $15_{-8}^{+0**}$ km & $6_{-3}^{+7}$ km & $7_{-3}^{+5}$ km & $7_{-3}^{+5}$ km \\[0.1cm]
    & Flux$_{pl, \Gamma = 2}$ & $(2 \pm 1) \times 10^{-15}$ & $3_{-1}^{+2} \times 10^{-15}$ & $4_{-1}^{+2} \times 10^{-15}$ s$^{-1}$ & $(3 \pm 1) \times 10^{-15}$ & $(3 \pm 1) \times 10^{-15}$ & $4_{-2}^{+1} \times 10^{-15}$ \\[0.1cm]
    & (ergs cm$^{-2}$ s$^{-1}$) & & & &  & &\\[0.1cm]
    & C-statistic & $24.04$ & $19.18$ & $21.94$ & $18..24$ & $23.59$ & $20.39$ \\[0.1cm]
    & $\chi^2$/d.o.f & $25.03/15$ & $18.36/15$ & $21.30/15$ & $17.30/15$ & $23.83/15$ & $19.85/15$ \\[0.1cm] \hline
    \multirow{4}{1.5in}[0.0cm]{\centering \textsc{wabs*tbabs*(nsmaxg + pegpwrlw) \\(H atmosphere, $B=10^{13}$G)}} & $f$ & $1.5_{-0.2}^{+0.3}$ &	$1.4_{-0.1}^{+0.2}$ & $1.4_{-0.2}^{+0.3}$ & $0.7 \pm 0.1$ & $1.2 \pm 0.2$ & $1.0 \pm 0.2$ \\[0.1cm]
    & $N_{H, SMC}$ & $1_{-1}^{+3*} \times 10^{21}$ cm$^{-2}$& $2.5_{-2.3}^{+3.8} \times 10^{21}$ cm$^{-2}$ & $2.3_{-1.8}^{+2.7} \times 10^{21}$ cm$^{-2}$& $1_{-1}^{+3*} \times 10^{21}$ cm$^{-2}$ & $< 2^{*} \times 10^{21}$ cm$^{-2}$ & $(2 \pm 2^*) \times 10^{21}$ cm$^{-2}$ \\[0.1cm]
    & $T_{eff}$ & $1.7_{-0.1}^{+0.2} \times 10^6$ K & $1.74_{-0.08}^{+0.20} \times 10^6$ K & $(1.8 \pm 0.1) \times 10^6$ K & $1.7_{-0.1}^{+0.2} \times 10^6$ K & $1.75_{-0.07}^{+0.15} \times 10^6$ K & $1.8_{-0.1}^{+0.2} \times 10^6$ K \\[0.1cm]
    & $(R_{em}/R_{NS})$ & $1_{-0.3}^{+0.0**}$ & $1_{-0.3}^{+0.0**}$ & $1_{-0.2}^{+0.0**}$ & $1_{-0.3}^{0.0*}$  & $1_{-0.2}^{+0.0**}$ & $1_{-0.2}^{+0.0**}$ \\[0.1cm]
    & Flux$_{pl, \Gamma = 2}$ & $(1 \pm 1^{*}) \times 10^{-15}$ & $1.5_{-1.2}^{+1.5} \times 10^{-15}$ & $1.8_{-1.3}^{+1.4} \times 10^{-15}$ s$^{-1}$ & $1.6_{-1.2}^{+1.4} \times 10^{-15}$ & $(1 \pm 1^{*}) \times 10^{-15}$ & $1.7_{-1.3}^{+1.5} \times 10^{-15}$ \\[0.1cm]
    & (ergs cm$^{-2}$ s$^{-1}$) & & & &  & &\\[0.1cm]
    & C-statistic & $29.03$ & $26.47$ & $38.71$ & $23.19$ & $32.80$ & $28.48$ \\[0.1cm]
    & $\chi^2$/d.o.f & $31.56/15$ & $27.48/16$ & $41.57/16$ & $23.18/15$& $36.32/16$ & $30.04/16$ \\[0.1cm] \hline
    \multirow{4}{1.5in}[0.0cm]{\centering \textsc{wabs*tbabs*nsmaxg\\(C atmosphere, $B=10^{12}$G)}} & $f$ & $1.6_{-0.4}^{+0.2}$ &	$1.4 \pm 0.2$ & $1.5 \pm 0.2$ & $0.9_{-0.3}^{+0.1}$ & $1.2 \pm 0.2$ & $1.0_{-0.1}^{+0.2}$ \\[0.1cm]
    & $N_{H, SMC}$ & $3_{-3}^{+8*} \times 10^{21}$ cm$^{-2}$ & $9_{-7}^{+9} \times 10^{21}$ cm$^{-2}$ & $7_{-4}^{+7} \times 10^{21}$ cm$^{-2}$ & $2.11_{-2.06}^{+0.55} \times 10^{22}$ cm$^{-2}$ & $3_{-3}^{+7*} \times 10^{21}$ cm$^{-2}$ & $6_{-4}^{+6} \times 10^{21}$ cm$^{-2}$ \\[0.1cm]
    & $T_{eff}$ & $(3.3 \pm 0.4) \times 10^6$ K & $3.0_{-0.3}^{+0.4} \times 10^6$ K & $3.1_{-0.2}^{+0.4} \times 10^{6}$ K & $2.55_{-0.09}^{+0.88} \times 10^6$ K & $(3.2 \pm 0.3) \times 10^6$ K & $3.2_{-0.2}^{+0.3} \times 10^6$ K \\[0.1cm]
    & $(R_{em}/R_{NS})$ & $0.3_{-0.1}^{+0.2}$ & $0.4_{-0.1}^{+0.5}$ & $0.4_{-0.1}^{+0.3}$ & $1.0_{-0.7}^{+0.0**}$  & $0.3_{-0.1}^{+0.3}$ & $0.4_{-0.1}^{+0.2}$ \\[0.1cm]
    & C-statistic & $31.37$ & $18.03$ & $26.03$ & $17.38$ & $26.57$ & $25.12$ \\[0.1cm]
    & $\chi^2$/d.o.f & $30.99/16$ & $17.89/16$ & $25.58/16$ & $16.88/16$& $27.10/16$ & $24.36/16$ \\[0.1cm] \hline
\end{tabular}
\justify
Note: We use the same notations as in Table~\ref{table:spec_e0102}. Refer Fig.~\ref{fig:regions} for the locations of the different background regions considered. Though the continuum thermal emission, position, strength and width of emission lines is similar within the  error limits, the central values of these parameters can vary for different background regions. Since the thermal emission and the lines are fixed while fitting the source, these differences lead to varying goodness of the fits. For example, for ``Region 1" the best fitting width of Mg line is $0_{-0}^{+0.8}$ as compared to $0.07 \pm 0.03$ for the combined spectrum. Using zero width for the Mg emission line complex when fitting the source leads to residuals at $\sim 1.3$ keV, and thus bad fits. However, allowing all the background parameters to vary would increase the degrees of freedom reduce our confidence in the best fitting models.
\end{table}
\end{landscape}
}

\subsection{Search for X-ray pulsations}
\label{sec:pulse}
\begin{figure}
    \centering
    \includegraphics[width=0.45\textwidth]{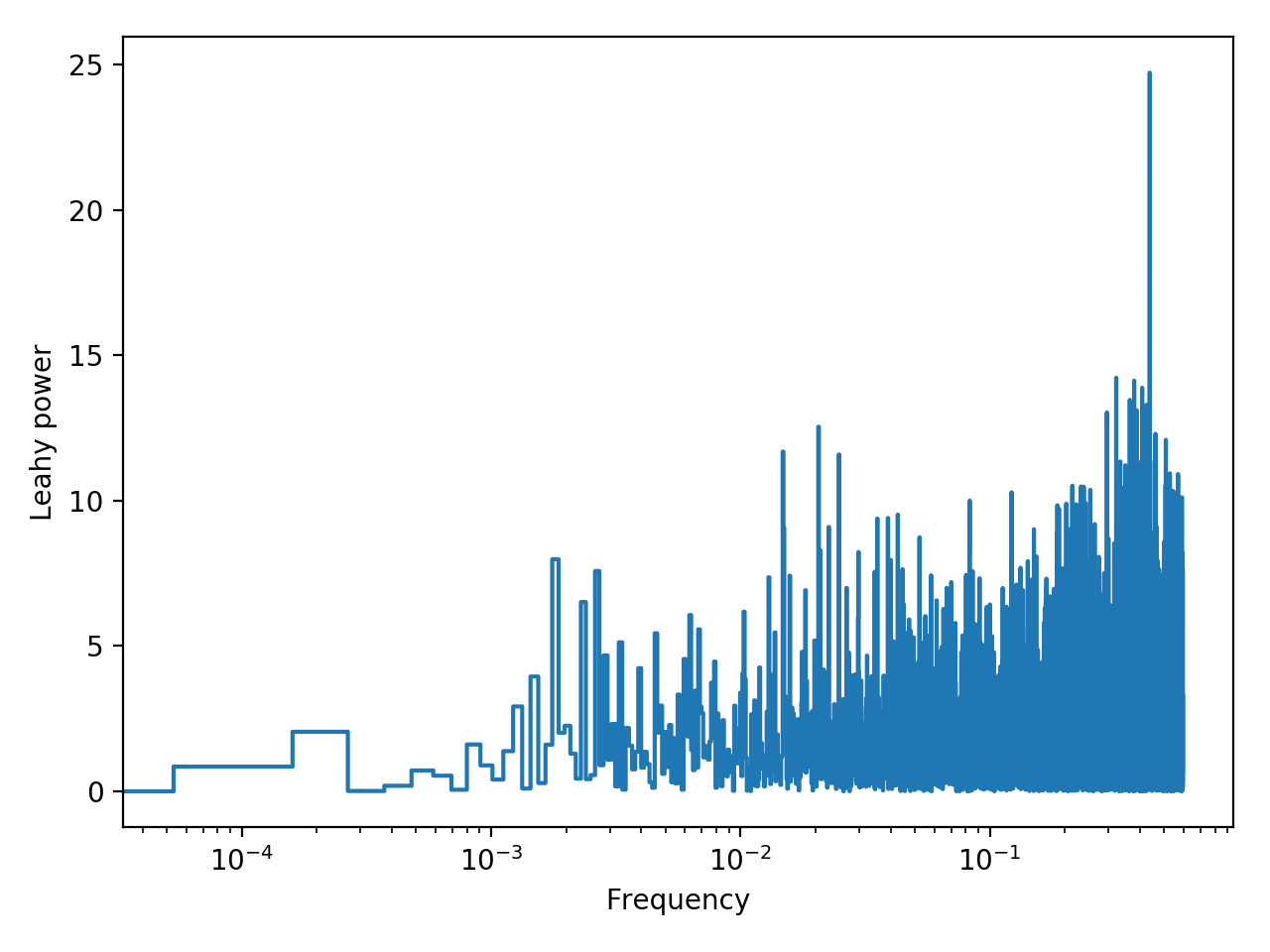}
    \caption{Leahy normalised power-spectrum of ObsID 6765 showing a peak at 0.44 Hz. The single trial probability of having a Leahy normalized power $\geq 24.7$ is $4.27 \times 10^{-6}$. With $5573$ frequency bins in this power spectra, this corresponds to a false alarm probability of $0.024$.}
    \label{fig:pulse}
\end{figure}

Pulsations in the X-ray light curve would reveal the presence of hotspots and their geometry. We check for periodicity in individual observations where the time difference between successive frames (i.e. the time resolution) is $\Delta t < 1$~s. Since the X-ray spectrum indicates a dominant single thermal component, we do not expect the pulse (if any) to have significant harmonics. In addition, the $\sim 1$ s time resolution of the ACIS observations used does not allow us to identify harmonics within a single time period (note that even magnetars have spin period of only a few seconds). For 13 observations, $\Delta t = 0.84$ s, allowing us to probe frequencies up to $0.6$ Hz. We extract barycentre-corrected light curves from each of these observations with the smallest allowed bin time (=$\Delta t$) using the CIAO 4.10 tools  \texttt{axbary} and \texttt{dmextract}. We then construct power spectra using the tool \texttt{apowerspectrum} of CIAO 4.10 and normalise them according to \citet{leahy1983}.  Leahy-normalised power-spectra follow a $\chi^2$ distribution with two degrees of freedom. We find a strong pulsation candidate at $0.44$ Hz ($P = 2.28$ s) for ObsID 6765 (Fig.~\ref{fig:pulse}). This frequency has a Leahy-normalised power of $24.7$. The single-trial probability of having Leahy-normalised power $ > 24.7$ is $4.27 \times 10^{-6}$ ($\sim 4.6\sigma$). Given that this power spectrum has $5573$ frequency bins, this corresponds to a false alarm probability of $0.024$. However, we do not find any similarly strong signal in the remaining 12 power spectra (i.e all signals in the remaining 12 power spectra have false alarm probability $> 0.35$, indicating that these are likely  due to Poisson noise). Considering searches over all these power spectra, the false-alarm probability of the 0.44 Hz pulsation candidate rises to 0.31. We also ran a $Z_n^2$ test on the ObsID 6765 using the phase calculated with a constant period $P = 2.28$s, but do not find clear rotational variability (significance of variability $< 2\sigma)$. A single long-exposure observation of this source would allow  a deeper search for pulsations.

\section{Discussion}
\label{sec:disc}
\subsection{Absorption} 
Most of our fits prefer a larger inferred absorption column (up to $\sim 10^{22}$ cm$^{-2}$ in some cases) than observed for SNR 1E0102.2-7219  ($5.76\times10^{20}$ cm$^{-2}$ of SMC absorption, plus $5.36\times10^{20}$ cm$^{-2}$ of Galactic absorption \citealt{plucinsky2017}, consistent with the more recent analysis of \citealt{alan2018}, which gives $N_{H, Gal} = 4.5 \times 10^{20}$ cm$^{-2}$, 
$N_{H, SMC} = 8 \times 10^{20}$ cm$^{-2}$). Recent calculations of the likely internal absorption column to the central NS produced by a $\sim$ 2000-year-old SNR are much smaller \citep{Alp18}. However, we identify three caveats to this apparent discrepancy. First, the constraints on $N_H$ are generally quite weak, such that the observed SNR $N_H$ cannot be ruled out. Second, any contribution by SNR ejecta to the absorption will have much higher abundances than the SMC in general. Third, the low-ionisation gaseous ring (shell?) around the NS discovered by \citet{vogt2018} may be thick enough to provide substantial extinction, though its nature and density have not yet been quantified. Such an optical feature/nebula has only been suggested around one other CCO, CXOU J085201.4-461753 in SNR G266.2-1.2/Vela Jr.  \citep{pavlov2001c,pellizzoni2002,migani2019}. However, whether this optical nebula is even associated with the CCO in Vela Jr. is not certain.

\subsection{Atmosphere} 
We found significant residuals when fitting the NS with any hydrogen atmosphere model. However, we found significantly better fits when using a ($10^{12}$ G) C atmosphere, or a blackbody plus power-law.  This suggests that either this NS hosts a C atmosphere, or that the atmosphere is described by some combination of composition, depth (e.g. an optically thin H atmosphere, which could have accumulated over 2000 years), and/or magnetic field, which we have not tried, and which might be reasonably represented by a blackbody. (Although we have tried a number of models, we cannot claim to have exhausted all the possibilities, especially if the surface is not homogeneous.)

A (non-magnetic) carbon atmosphere spectrum is a good description of the spectrum of the youngest known NS, the CCO in the Cassiopeia~A SNR \citep{ho09}.  Similarly, a carbon atmosphere spectrum can fit the spectrum of three other CCOs (see Section~\ref{sec:nature}), two with ages of 1--2~kyr and one much older at 27~kyr \citep{klochkov2013,Klochkov16,doroshenko2018}.  As shown recently by \citet{Wijngaarden19}, a carbon atmosphere can be present on a NS of sufficient youth ($\lesssim1000\mbox{ yr}$), as its high temperature burns any surface hydrogen or helium.  After this age, the temperature becomes low enough to allow accumulation of hydrogen even at very low accretion rates from the interstellar medium and thus formation of a hydrogen atmosphere.  With an age of $2050\pm600\mbox{ yr}$, E0102 may be near the transition between a carbon atmosphere and a thin hydrogen atmosphere, when the latter approaches an optical thickness $\tau_\nu\sim 1$, such as that which seems to exist on (much older) X-ray isolated NSs such as RX~J0720.4$-$3125 \citep{motchetal03} and RX~J1856.5$-$3754 \citep{hoetal07}.

In contrast to the above CCOs that are fit with a carbon atmosphere spectrum which assumes no or low magnetic fields, our best-fit carbon spectrum assumes $B=10^{12}\mbox{ G}$.  The spin and spectral properties of three other CCOs \citep{mereghettietal02,sanwaletal02,Halpern10,gotthelf13} indicate these three CCOs have $B\sim10^{10}-10^{11}\mbox{ G}$ (see Section~\ref{sec:nature}).  A low magnetic field currently could be due to a stronger field that was buried by initial fallback of supernova material and is only now emerging at the surface \citep{Ho11,ho15}.  The emergence timescale depends on the amount of material accreted, such that E0102 could have accreted less and thus its field has already emerged to values typical of pulsars.  Alternatively, E0102 could be a magnetar with a subsurface field that is $\gtrsim10^{14}\mbox{ G}$ and a surface field that is still emerging.  If E0102 is a magnetar, it may undergo a magnetar outburst in the future.  This is an interesting prospect given the extensive monitoring of SNR~1E~0102.2$-$7219 as a calibration source for telescopes such as {\it Chandra} and {\it NICER}.

\begin{figure}
    \centering
    \includegraphics[width = 0.45\textwidth]{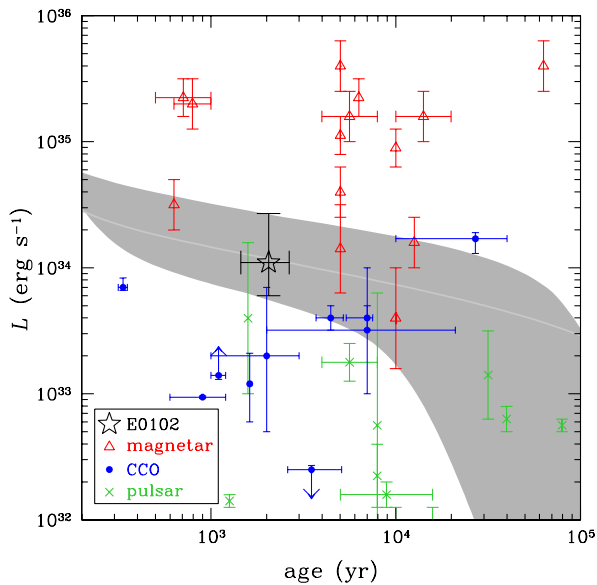}
    \caption{Thermal luminosity versus age of E0102 (black star) with respect to known pulsars (green crosses), CCOs (blue circles) and magnetars (red triangles).  The shaded region shows the cooling scenario for $M = 1.4$M$_{\odot}$, $R = 11.6$ km and a light element envelope. The upper bound of this region is obtained by only considering superconducting protons, while the lower bound is achieved including superfluidity of neutrons as well. The solid line is the same cooling scenario as the upper bound but with $M = 1.2$M$_{\odot}$ and an iron envelope. E0102 is brighter than CCOs and pulsars, but less luminous than magnetars  of similar age.}
    \label{fig:cooling}
\end{figure}

\subsection{Nature of the neutron star: \label{sec:nature}}

We also look into the general properties of CCOs, magnetars and young pulsars to further study the nature of E0102 and classify the NS.

\citet{deLuca08,deLuca17,Halpern10} summarise all known CCOs\footnote{see http://www.iasf-milano.inaf.it/{\texttildelow}deluca/cco/main.htm for the updated list.}. We see that most CCOs have thermal luminosities $\sim 10^{33}$ ergs s$^{-1})$.  XMMU J173203.3-344518 in SNR G353.6-0.7 has a thermal luminosity, $1.3 \times 10^{34}$ ergs s$^{-1}$, comparable with that of E0102, though it requires relatively extreme cooling parameters \citep{Klochkov15}. Its X-ray spectrum can be best fit using a two temperature blackbody \citep[$kT_1 \approx 0.4$ keV, $R_1 \approx 1.5$ km, $kT_2 \approx 0.6 - 0.9$ keV, $R_2 \approx 0.2 - 0.4$ km;][]{halpern10c} or a non-magnetic NS with a C atmosphere \citep[$kT \approx 0.19$ keV, $R \approx 13$ km;][]{klochkov2013}. The observed lack of pulsations supports the C atmosphere model where the entire NS surface emits radiation. Such a  non-magnetised carbon atmosphere, with $kT \sim 0.15$ keV and emission from the entire NS, has also been proposed for the CCOs CXOU J232327.9+584842 \citep[in the SNR Cas A,][]{ho09}, CXOU J160103.1-513353 \citep[in G330.2+1.0,][]{doroshenko2018} and CXOU J181852.0-150213 \citep[in G15.9+0.2,][]{Klochkov16}. X-ray emission of other CCOs can be adequately fit ($\chi^2_{\nu} \leq 1.1$) using a BB ($kT \sim 0.5$ keV) or a non-magnetic NS atmosphere (NSA) model ($kT \sim 0.3$ keV) \citep{gotthelf13, Halpern10,lovchinsky11}. However, a second BB ($kT_1 = 0.2 - 0.4$ keV, $R_1 =  2 - 4 km$, $kT_2 = 0.5 - 0.9$ keV, $R_2 \lesssim 1$km) or NSA component often improves the fit.
 
Timing solutions of CCOs showing rotational variability (PSR J0821-4300, PSR 1852+0040 and 1E 1207.4-5209)  reveal periods between 0.1 \& 0.4s, and surface $B < 10^{11}$ G, indicating that CCOs have relatively low $B$ fields \citep{gotthelf13, Halpern10}. The E0102 NS cannot be fit using a single BB or NSA model ($\chi_{\nu}^2 \sim 2$), and the C atmosphere fit needs a higher magnetic field ($B = 10^{12}$) and relatively high temperature ($kT \sim 0.3$ keV)  for a good fit. Although other CCOs show no indications of radio pulsations or synchrotron nebulae, it is not clear whether their observed spindowns correctly indicate their surface $B$ fields.  For example, the high pulsed fraction of the NS in Kes 79 strongly indicates a high surface $B$ field, possibly in a strongly multipolar configuration \citep{Bogdanov14}. It has been suggested that the low measured fields in CCOs are due to burial of the field by fallback \citep[e.g.][]{Ho11}, in which case a normal $B$ field for E0102 would not be truly unusual (though it would raise questions about why it has not shown radio pulsar behaviour).

The thermal emission of magnetars with $B \sim 10^{14} - 10^{15}$ G in their quiescent state is very similar to that of CCOs and generally cannot be distinguished by X-ray spectra alone. 
We use the McGill online magnetar catalogue\footnote{http://www.physics.mcgill.ca/~pulsar/magnetar/main.html} \citep{Olausen14} to study and compare the X-ray properties of magnetars. Magnetars of age $\mathcal{O}(10^{3}$ yrs $)$ have thermal X-ray luminosities, $L_{2 - 10 keV} \sim 10^{33} - 10^{35}$ ergs s$^{-1}$ during their quiescent state, nicely encompassing the E01012 NS's luminosity (see Fig.\ref{fig:cooling}). Like CCOs, the X-ray spectra of magnetars can be fit using two BB components, or a BB+PL. The observable differences between magnetars and CCOs comes from X-ray variability --- magnetars typically show fast, bright outbursts, and/or show pulsations (with periods $\sim 2 - 12$ s) revealing rapid spindown indicative of high $B$ fields. Our
 tentative 2.28 s periodic signal is near the lower limit of the rotation period for known magnetars. We did not identify any long-term X-ray variability from E0102 over the years 2003-2017. From \citet{vigano2013}, we see that the expected outburst rate for magnetars of age 2000 years is $\sim 0.05$/yr. Thus the non-detection of an X-ray outburst doesn't rule out the possibility that E0102 could be a magnetar in quiescence.

A thermal component has been detected in very few young (ages below $10^4$ years) pulsars with ``normal" $B$-field strengths, $B\sim 10^{12}$ G, due in part to the bright non-thermal pulsed and pulsar wind X-ray emission. 
Only four radio pulsars have measured thermal X-ray spectral components (each also has nonthermal components) and inferred ages $<10^4$ years; these are PSR J1119-6127 \citep[][$\tau$=1600 years]{Gonzalez05}, PSR J1357-6429 \citep[][$\tau$=7300 years]{Zavlin07},  
PSR J1734-3333 \citep[][$\tau$=8100 years]{Olausen13}, and 
PSR B1509-58 \citep[][$\tau$=1700 years]{Hu17}. 
Interestingly, these four young pulsars all show high magnetic fields; estimating $B$ from $P$ and $\dot{P}$, $B=4\times10^{13}$ G, $8\times10^{12}$ G, $5\times10^{13}$ G, and $1.5\times10^{13}$ G respectively. Blackbody fits to their thermal components give $T=2.4_{-0.2}^{+0.3}\times10^6$ K, $R=3.4_{-0.3}^{+1.8}$ km, $L_{X,bb}=2_{-0.4}^{+2.5}\times10^{33}$ ergs s$^{-1}$ for J1119, 
$T=(1.7\pm 0.2)\times10^6$ K, $R=2.5 \pm 0.5$ km, $L_{X,bb} \sim 2\times10^{32}$ ergs s$^{-1}$ for J1357, 
$T=(3.5 \pm 0.7)\times10^6$ K, $R=0.45_{-0.20}^{+0.55}$ km, $L_{X,bb} \sim 2\times10^{32}$ ergs s$^{-1}$ for J1734, 
and $T=(1.7 \pm 0.1)\times10^6$ K, $R\sim9$ km, $L_{X,bb}\sim 9\times10^{33}$ ergs s$^{-1}$ for B1509. Unusually, the thermal components show extremely strong pulsations in two of these (48--74\% pulsed fraction for J1119 \citealt{Gonzalez05,Ng12}, $>$50\% pulsed fraction for J1357; J1734 has only an upper limit on the pulsed fraction of $<$60\%, and pulsation searches on B1509's thermal component were not possible due to the strong pulsed nonthermal emission). Such strong pulsations cannot be achieved for any hotspot geometry without substantial radiative beaming, which requires higher $B$ fields $>10^{14}$ G \citep{Bogdanov14}.
Indeed, J1119 (estimated $B=8\times10^{12}$ G) underwent a series of transient magnetar bursts \citep{Archibald16}, strongly indicating that J1119, and possibly the others, have magnetar-strength internal $B$ fields \citep[e.g.][]{Ho11,ho15,Vigano12,Bernal13}. The E0102 NS has a higher thermal luminosity than these high-B young pulsars, and an apparently larger emitting radius. It appears that if  E0102 NS is a radio pulsar, it is likely to resemble these high-B pulsars, and may be a hidden magnetar. 

\begin{figure}
    \centering
    \includegraphics[width=0.45\textwidth]{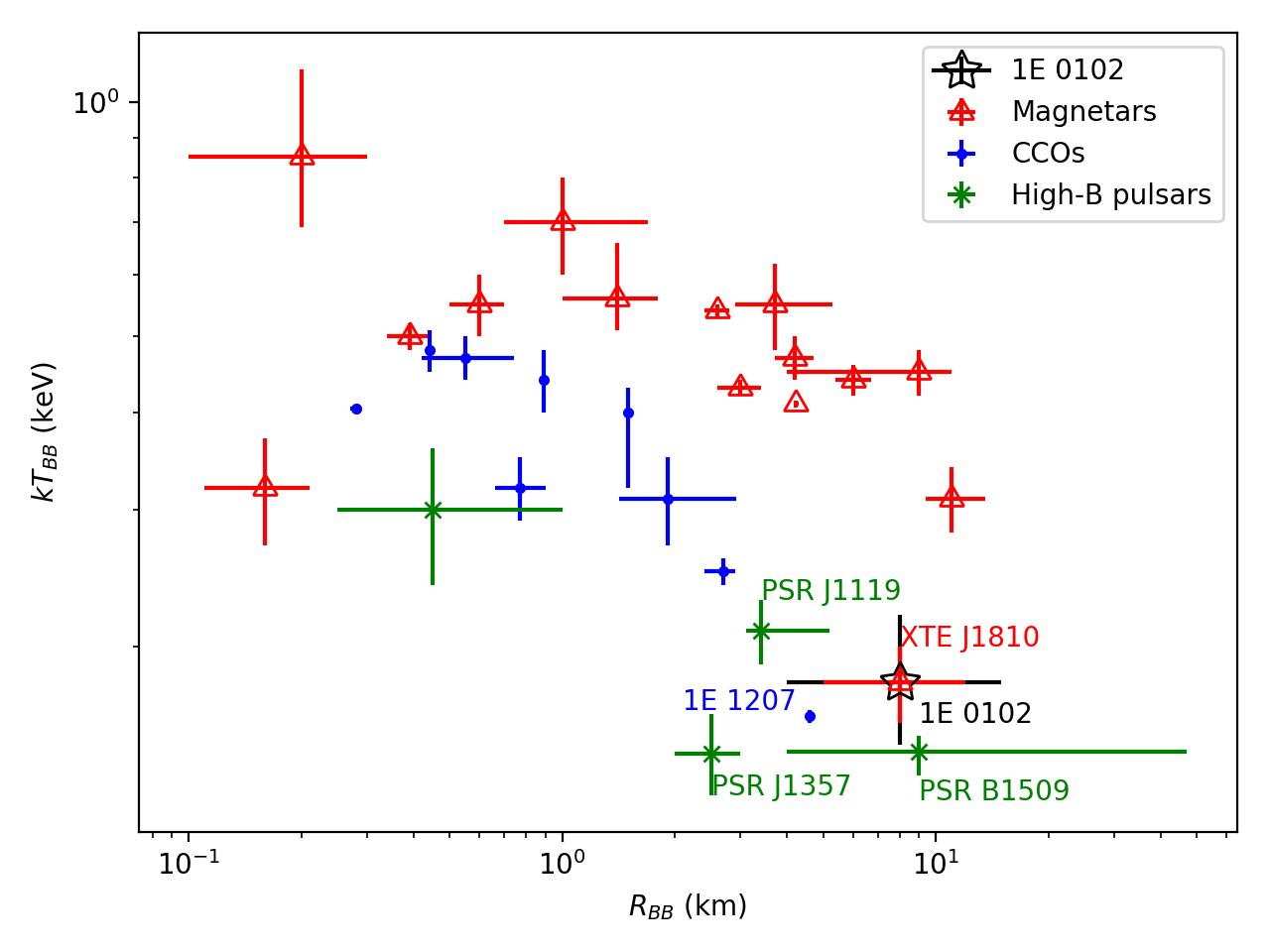}
    \caption{Comparison of $kT_{BB}$ and $R_{BB}$ on fitting a BB model to different classes of NS. Magnetars are usually hotter and brighter than CCOs and high-B pulsars. The leftmost magnetar ($R_{BB} \sim 0.2$~km, $kT\sim 0.3$~keV is a low magnetic field ($B = 6\times 10^{12}$~G) faint ($L_{bol} \sim 4 \times 10^{31}$~ergs~s$^{-1}$) system, SGR 0418+5729 \citep{rea2013}.     The position of E0102 is consistent with that of some CCOs, high-B pulsars, and magnetars.}.
    \label{fig:hotspot}
\end{figure}

In Figure~\ref{fig:cooling}, we show the luminosity and age of E0102, as well as those of CCOs (see \citealt{reynoldsetal06,Klochkov16}, and references in \citealt{luo2015}) and magnetars and rotation-powered pulsars \citep{changetal12,Olausen13,vigano2013,Hu17}.  The shaded region indicates luminosity as a function of age for theoretical models of NS cooling, assuming a $1.4M_\odot$ NS built using the APR equation of state and a light element envelope. The upper boundary considers slower cooling due to superconducting protons, while the lower boundary considers more rapid cooling due to Cooper pair formation and breaking of superfluid neutrons. Meanwhile, the solid line is for the same model as the upper boundary, but with an iron envelope and $1.2M_\odot$ NS (see \citealt{luo2015}, and references therein, for details).

Regardless of the atmosphere model we use, we find a very high thermal luminosity for the E0102 NS, which might be explained either through slow loss of heat from the supernova explosion, or decay of an initially strong magnetic field.
Our carbon atmosphere NS spectral model  gives a  bolometric thermal luminosity $L_{bol} = 1.1_{-0.5}^{+1.6} \times 10^{34}$ ergs s$^{-1}$. Comparing E0102's parameters to other known young NSs, we are struck by E0102's relatively high inferred temperature and thermal luminosity. E0102's thermal luminosity is larger than most rotation-powered pulsars. Only young CCOs like PSR J1852+0040 (in the SNR G33.6+0.1), PSR J0821-4300 (in the SNR Puppis A) and PSR J1210-5226 (in the SNR G296.5+10.0), the young high-$B$ pulsars B1509-58 and J1119-6127, and magnetars have thermal emission within the error limits of E0102's luminosity.

We compare the emitting radius and temperature of the observed blackbody radiation from CCOs, high-$B$ pulsars, and magnetars in Fig.~10. Unfortunately, this does not clearly distinguish between the three groups of NSs--although E0102's properties are somewhat unusual for each class, there are members of each class with similar properties.

\section{Summary and conclusions}

In this work, we used the combined spectra from the {\it Chandra} ACIS-S observations of SNR 1E 0102-7219 to argue that the compact object detected by \citet{vogt2018} in the SNR is a neutron star and to constrain its nature and properties. The observed X-ray spectrum of this source cannot be explained as a concentration of SNR emission, clearly requires an additional soft blackbody-like source, and confirms the detection  by \citet{vogt2018} of a neutron star in this SNR. The emission at higher energies cannot be modelled by a simple BB and requires an additional non-thermal power-law component. Among the NS models, we see that H atmosphere models at any $B$-field strength fail to fully model the X-ray spectra. Adding a power-law component successfully models the emission at higher energies, but the residuals around $\sim 1$ keV are still not explained. A $10^{12}$ G carbon atmosphere NS model better fits these residuals, as well as emission at the higher energies.

The best fit temperature and luminosity of this compact object in E0102 is higher than most NSs observed. Comparing the thermal luminosity to other pulsars indicates that it is unlikely that this compact object is powered by rotation. The high temperature and the presence of hotspots suggest that this source is powered by its high magnetic fields, like the magnetars and some high-B radio pulsars. However, there are a few observed low-B NSs with inferred ages $ < 10^4$ years and similar temperatures, so the emission we see may be heat from the supernova explosion that formed this NS. Identifying and studying young neutron stars like this one is essential to understand the physical mechanisms responsible for the high thermal luminosity.

The greatest challenge in studying this NS is the bright background emission from the SNR itself. Though the parameter values do not change significantly when using different background models, the C-statistic and the $\chi^2$ value indicate the quality of the fits change significantly. The proposed Lynx \citep{Lynx18} X-ray mission, with higher effective area,  greatly improved spectral resolution, and similar angular resolution as Chandra, would permit a more powerful analysis of such compact objects in bright SNRs. Lynx's high spectral resolution  would allow us to identify and filter out the background emission lines from the source spectra. It is also possible that Lynx could measure spectral features (e.g. edges) in NS spectra, identifying the nature of the NS atmosphere. Lynx's higher effective area could also  detect X-ray pulsations (if present), permitting the measurement of $P$ and $\dot{P}$, and thus constraining the magnetic field strength. The expansion of the study of NS surfaces to nearby galaxies would add greatly to our understanding of NSs. 

\section*{Acknowledgements}

We thank E. Bartlett and F. Vogt for discussions. We also thank the referee for a thoughtful report and suggestions. COH is supported by NSERC Discovery Grant RGPIN-2016-0460 and a Discovery Accelerator Supplement.  WCGH acknowledges support from the Science and Technology Facilities Council through grant number ST/M000931/1.






\bibliographystyle{mnras}
\bibliography{E0102}







\bsp	
\label{lastpage}
\end{document}